\documentclass{aastex631}

\usepackage{graphicx}   

\usepackage{multirow}
\usepackage{array}

\usepackage{amsmath}
\usepackage[percent]{overpic}
\usepackage{array}      
\usepackage{verbatim}   
\usepackage{bm}

\usepackage{graphbox}

\usepackage{mwe}
\usepackage{rotating}
\usepackage{enumerate}
\usepackage{color}
\usepackage{epsfig}     
\usepackage{url}        
\usepackage{subfigure}  
\usepackage{multirow} 
\usepackage{wrapfig}
\usepackage{float}
\usepackage[hyphens]{url}
\usepackage{hyperref}
\usepackage{color}

\begin{document}

\title{Coronal Models and Detection of Open Magnetic Field}

\correspondingauthor{Eleanna Asvestari}
\email{eleanna.asvestari@helsinki.fi}

\author[0000-0002-6998-7224]{Eleanna Asvestari}
\affiliation{Department of Physics, University of Helsinki, P.O. Box 64, 00014, Helsinki, Finland}

\author[0000-0003-4867-7558]{Manuela Temmer}
\affiliation{Institute of Physics, University of Graz, Universitätsplatz 5, 8010 Graz, Austria}

\author[0000-0002-2633-4290]{Ronald M. Caplan}
\affiliation{Predictive Science Inc., 9990 Mesa Rim Road, Suite 170, San Diego, CA 92121, USA}

\author[0000-0003-1662-3328]{Jon A. Linker}
\affiliation{Predictive Science Inc., 9990 Mesa Rim Road, Suite 170, San Diego, CA 92121, USA}

\author[0000-0002-2655-2108]{Stephan G. Heinemann}
\affiliation{Max-Planck-Institut für Sonnensystemforschung, Justus-von-Liebig-Weg 3, 37077 G\"ottingen, Germany}
\affiliation{Institute of Physics, University of Graz, Universitätsplatz 5, 8010 Graz, Austria}

\author[0000-0001-8247-7168]{Rui F. Pinto}
\affiliation{LDE3, DAp/AIM, CEA Saclay, 91191 Gif-sur-Yvette, France}
\affiliation{IRAP, Université de Toulouse; UPS-OMP, CNRS; 9 Av. colonel Roche, BP 44346, F-31028 Toulouse cedex 4, France}

\author[0000-0002-6038-6369]{Carl J. Henney}
\affiliation{Air Force Research Laboratory, Space Vehicles Directorate, KAFB, NM, USA}

\author[0000-0003-1662-3328]{Charles N. Arge}
\affiliation{Heliophysics Science Division, NASA Goddard Space Flight Center, Code 671, Greenbelt, MD,
20771, USA}

\author[0000-0003-2061-2453]{Mathew J. Owens}
\affiliation{Department of Meteorology, University of Reading, Earley Gate, P.O. Box 243, Reading RG6 6BB, UK}

\author[0000-0001-9806-2485]{Maria S. Madjarska}
\affiliation{Max-Planck-Institut für Sonnensystemforschung, Justus-von-Liebig-Weg 3, 37077 G\"ottingen, Germany}
\affiliation{Space Research and Technology Institute, Bulgarian Academy of Sciences, Acad. Georgy Bonchev Str., Bl. 1, 1113, Sofia, Bulgaria}

\author[0000-0003-1175-7124]{Jens Pomoell}
\affiliation{Department of Physics, University of Helsinki, P.O. Box 64, 00014, Helsinki, Finland}

\author[0000-0001-7662-1960]{Stefan J. Hofmeister} 
\affiliation{Columbia Astrophysics Laboratory, Columbia University, 550 West 120th Street, New York, NY 10027, USA}
\affiliation{Leibniz-Institute for Astrophysics Potsdam, An der Sternwarte 16, 14482 Potsdam, Germany}

\author[0000-0002-5681-0526]{Camilla Scolini}
\affiliation{Institute for the Study of Earth, Oceans, and Space, University of New Hampshire, Durham, NH 03824, USA}

\author[0000-0002-7676-9364]{Evangelia Samara}
\affiliation{Heliophysics Science Division, NASA Goddard Space Flight Center, Code 671, Greenbelt, MD,
20771, USA}

\received{13-Sep-2023}
\submitjournal{Astrophysical Journal}
\shorttitle{Coronal models and open field}
\shortauthors{Asvestari et al.}

\begin{abstract}

A plethora of coronal models, from empirical to more complex magnetohydrodynamic (MHD) ones, are being used for reconstructing the coronal magnetic field topology and estimating the open magnetic flux. However, no individual solution fully agrees with coronal hole observations and in situ measurements of open flux at 1~AU, as there is a strong deficit between model and observations contributing to the known problem of the missing open flux. In this paper we investigate the possible origin of the discrepancy between modeled and observed magnetic field topology by assessing the effect on the simulation output by the choice of the input boundary conditions and the simulation set up, including the choice of numerical schemes and the parameter initialization. In the frame of this work, we considered four potential field source surface based models and one fully MHD model, different types of global magnetic field maps and model initiation parameters. After assessing the model outputs using a variety of metrics, we conclude that they are highly comparable regardless of the differences set at initiation. When comparing all models to coronal hole boundaries extracted by extreme ultraviolet (EUV) filtergrams we find that they do not compare well. This miss-match between observed and modeled regions of open field is a candidate contributing to the open flux problem.

\end{abstract}


\section{Introduction} \label{sec:intro}

A well-known conundrum in space physics is the ``open flux problem'' according to which open flux estimates derived from in situ spacecraft measurements at 1~au \citep[see for example][]{Lockwood_1_2009, Lockwood_2_2009, Owens_2018, Frostetal22} are systematically larger than estimates from: (1) calculations based on remote sensing observations of coronal holes (CHs) and the underlying magnetic flux and (2) model reconstructions of open field areas associated to CHs on the Sun \citep[see for example][and referencees therein]{linker2017, wallace19}. The problem rises from the fact that open flux is considered to originate from magnetic field regions predominantly associated with CHs \citep[][]{levine77, wang96, schwadron2005, fisk2006, schwenn06a, schwenn06b, antiochos07, cranmer2017, Hofmeister_etal_2017, Hofmeister_etal_2019, heinemann18}. This implies that the aforementioned systematic discrepancies should not occur. The scientific community has presently reached a consensus that open flux may also originate from sources other than CHs, resulting in an excess of open flux not accounted for so far in the calculations. To understand the exact magnitude of the missing open flux one needs to quantify the uncertainties of the coronal models used for reconstructing open fields, of the methodologies used to estimate open flux from observations, and of the observations themselves.  

In \cite{linker21}, to which we refer to hereafter as Paper~I, we characterized the uncertainties of EUV CH detection methods that are often used for deriving the open magnetic flux. We derived uncertainties in the open flux calculation of the order of about 26\% due to CH boundary variations when applying the different EUV extraction methods. By using magnetograms from different sources, ground-based and satellites with various processing levels, that uncertainty even doubled to about 45\%. From an MHD simulation, we also found that some open flux would be missing due to originating from regions that do not appear dark in EUV. In this paper we focus on determining model uncertainties in reconstructing open field areas associated to CHs. We also consider the effect on the simulation output of the ambiguity introduced by certain observational limitations affecting the boundary conditions used to initiate model simulations.

Current efforts in global coronal magnetic field modeling range from simplified empirical models to advanced magnetohydrodynamic (MHD) ones \citep[see for example][]{altschuler69, mikicetal99, linkeretal1999, lionelloetal1998, lionello01, rileyetal01, arge00, grothetal00, cohenetal07, wiegelmann_etal07, vanderholstetal10, vanderholstetal14, antiochos11, evansetal12, jiangetal12, guoetal16, pinto17, sachdevaetal19, sachdevaetal21, feng2020, wang_etal20, perri2022}. However, due to its simplicity and computational inexpensiveness, the Potential Field Source Surface (PFSS) model is the most commonly employed one. It forms one of the building blocks of the Wang-Sheeley-Arge \citep[WSA;][]{arge00} and of other similar semi-empirical models currently used in state-of-the-art space weather models, such as Enlil \citep[][]{odstrcil03, odstrcil05}, European Heliospheric FORecasting Information Asset \citep[EUHFORIA;][]{pomoell18}, and the multiple flux-tube solar wind model \citep[MULTI--VP;][]{pinto17}. All coronal models come with different layers of complexity and uncertainty, either of which could potentially account for the missing open flux. Sources of complexity and uncertainty can be grouped into two main categories: (1) those introduced by the choice of input boundary conditions, more precisely, the selected global magnetic field map, and (2) those related to the simulation set-up, that includes employed numerical schemes and the choice of parameter initialization.

In order to initiate a coronal model simulation, one needs to know the magnetic flux distribution at the inner boundary of the modeling domain. Traditionally, line-of-sight or vector magnetograms of the photospheric magnetic field are used; however, constructing appropriate datasets is challenging due to a number of limitations in current observational capabilities. Thus far, photospheric magnetic field measurements are routinely collected only from the Earth's vantage point and, as a result, at any given time we observe with adequate accuracy only a small sector of the solar disk that is within our field of view. For the rest of the solar surface, we either do not have observations at the same time, or they are subjected to strong projection effects towards the limb. Consequently, the reconstructed magnetograms, representing the radial magnetic field component of the entire photosphere, are in part based on outdated or on later date observations of the different sectors of the Sun as they progressively enter the Earth's field of view. This can lead to a lack of or outdated information on the magnetic field associated with CHs and active regions (ARs). An additional issue introduced is the consistent lack of accurate polar field observations, due to the inclination of the solar axis with respect to the ecliptic plane at a given period of the Earth's orbit and due to projection effects. Considering the importance of the open magnetic field from polar CHs in the global coronal magnetic field topology and subsequently the open flux estimates, it is expected that this issue can cause major discrepancies in our modeling efforts. Attempts in understanding the magnetic field topology of polar CHs are based on validating models against (1) white light and EUV observations, as well as in situ interplanetary measurements \citep[see for example][and references therein]{sunetal11, rileyetal19, schonfeldetal22, wagner22}, or (2) one-off past observational missions \citep[see for example the results from the SUNRISE mission in][]{prabhuetal20}. In the near future, validation efforts will also consider observations from the Solar Orbiter mission, which will progressively orbit at a higher inclination and close to the Sun \citep[][]{solankietal20, Harraetal21}. Nevertheless, at the time of writing, we are still far from producing magnetograms with accurate polar magnetic field values.

Beyond our observational limitations, full-Sun magnetograms developed based on magnetic field measurements obtained by different observatories, using different instruments and being constructed following different techniques, show qualitative differences \citep{riley2014, lietal2021, wang22}. To overcome this issue, normalization factors were introduced aiming in better connecting the magnetograms to each other. It has long been debated whether the differences among the magnetograms indicate uncertainties in the magnetic field measurements, regardless of the observatory, thus being the source of the known "open flux problem" \citep[see for example][and references therein]{riley2014, linker2017}. A recent study by \citet{wang22} showed that magnetic field corrections applied to Mount Wilson Observatory (MWO) or Wilcox Solar Observatory (WSO) data could solve the open flux issue, making the derived open fluxes consistent with the observed interplanetary magnetic field (IMF) magnitude. However, this is a temporary solution which does not resolve the problem since it solely adjusts the magnetogram values in order to obtain a result that is in agreement to in situ measurements. When it comes to using different magnetograms for coronal magnetic field modeling, it is essential to stress that a different input magnetogram will lead to a different model output \citep{Hayashi2016, linker2017, caplanetal2021, lietal2021}. All the aforementioned limitations introduce uncertainties to the reconstructed magnetograms. These uncertainties can be propagated to the simulation output. In our analysis, we focus only on the effect on the modeled topologies due to lack of or outdated information on ARs and the small variations in the magnetograms due to their generation method.

Apart from the boundary conditions discussed above, solving the mathematical models requires the use of a given numerical technique, and at the moment different algorithms are being used by the community, each having diverse characteristics. For example to solve the Laplace equation for magnetic field computations in the PFSS model, there are different numerical schemes that can be implemented, such as the spherical harmonics \citep[employed for example in][]{altschuler69, arge00, pomoell18} and the finite difference schemes \citep[employed for example in][]{caplanetal2021}. Both schemes require the initialization of different parameters, for example the number of spherical harmonics or the resolution of the grid, respectively, the selection of which can significantly affect the model output.
Similarly, when focusing on defining the modeled areas of the open field, the field line tracing method applied will have some influence on the final size of the modeled open field area. In our analysis we compare the output generated by the different numerical schemes and quantify the degree of agreement/disagreement between results. This is done by implementing different models as discussed later on.

In addition, the source surface is an important structure included in the PFSS model as it is the upper boundary of the simulation domain and defines whether a modeled field line will be closed or open. More precisely, all modeled field lines piercing through the source surface are labeled as open while those that bend back to the Sun below it are labeled as closed. Therefore, in the modeling domain, the height above the photosphere where the source surface is placed defines the size of the modeled areas of open field, and thus the size of the photospheric area where this open field is rooted at \citep[see Figure 1 in][]{asvestari19}. Considering that open field lines are primarily associated with CHs, this modeled photospheric area of open field would ideally map areas of CHs, which are observed in Extreme Ultra-Violet (EUV) filtergrams as dark regions. However, this is often not the case. The expansion of the coronal hole with height and the uncertainty of the exact depth we observe the structure in EUV may result in observed coronal hole areas that are not representative of those at the photosphere. This will thus lead to uncertainties when comparing observed to modeled areas. This hinders an additional potential issue of using photospheric data as boundary conditions to model the solar coronal structures. Moreover, the source surface height will also define the characteristics of the Heliospheric Current Sheet (HCS), a thin layer that separates the outward-pointing and inward-pointing open magnetic field lines.

Earlier research established that a typical value for the source surface height is that of 2.5 solar radii ($R_{\rm \odot}$) \citep[][]{altschuler69}. More recent studies have suggested or employed a much wider range of values (from 1.3$R_{\rm \odot}$ up to 3.6$R_{\rm \odot}$) \citep[][]{lee11, ardenetal14, asvestari19, asvestari20, wagner22}. They often show the need to use lower values in order to capture well the observed CHs in EUV \citep{asvestari19}, coronal magnetic field topologies observed during different solar cycles \citep[][]{lee11, ardenetal14}, or HCS crossings identified in near-Sun in situ measurements by the recent Parker Solar Probe mission \citep{badman20, panasenco20}. Another complexity of the source surface is its shape. While so far the majority of studies have assumed a spherical shape, non-spherical shapes have also been considered \citep[see for example][]{levineetal82, kruseetal20, kruseetal21}. The concept of a non-spherical source surface has already been suggested by the more complex MHD models \citep[][]{rileyetal06}. In our analysis, all models employed assume a spherical source surface, and the attention is placed on investigating the impact of the source surface height on the modeled magnetic field topologies.

With this International Space Science Institute (ISSI\footnote{http://www.issibern.ch/teams/magfluxsol/}) team, we brought together different coronal models, to investigate how the input boundary conditions and the simulation set up, including model parametrization and numerical scheme selection, affect the modeled areas of open magnetic field topology, contributing to the ``open flux problem''.  In Section \ref{sec:methods} we introduce briefly the coronal magnetic field models used in this study (Subsection \ref{sec:models}), the global magnetic field maps employed for initiating the simulation runs (Subsection \ref{sec:magnetograms}), the metrics we calculate for assessing the quality of the modeled open and closed field topology (Subsection~\ref{sec:metrics}), and the observational period and specific CH we simulate (Subsection \ref{sec:obs_CH}). More details on the models and the metrics used are given in Appendices \ref{ap1:models} and \ref{ap2:metrics} respectively. The results of our analysis are given in Section \ref{sec_results} and we discuss our key conclusions in Section \ref{sec_discussion}.

\section{Methodology and data}
\label{sec:methods}

\subsection{Global coronal models}
\label{sec:models}
In order to investigate how coronal magnetic field models compare to each other when they are based on the same set of physical assumptions but implemented using different methods we consider four PFSS-based models. They all employ different numerical solvers and initialization set up. The first model we employed is the WSA \citep[][]{arge00,arge03,arge04,mcgregor08,wallace19}, a combined empirical and physics-based model of the corona and solar wind. It is an improved version of the original Wang and Sheeley model \citep[][]{wang92,wang95}, coupling the PFSS and Schatten Current Sheet (SCS) model \citep[][]{schatten71}. In this study we consider only the computed magnetic field topology of WSA, and not the solar wind properties. The second model we used is the coronal model implemented in EUHFORIA. It is an adaptation of the WSA model \citep[][]{arge00}, with the \citet{mcgregor08} improvements for the nonphysical field line kinks between the PFSS and SCS model boundaries. Similar to WSA, from the EUHFORIA output we only considered the open and closed field maps produced. Both the WSA and EUHFORIA models employ a spherical harmonic method to solve the Laplace equation. The third model we adopted is the MULTI-VP which combines a method for reconstructing the three-dimensional open coronal magnetic field with a solar wind model that calculates the radial profiles and amplitudes of slow and fast solar wind flows. For this paper, MULTI-VP was setup with the PFSS+SCS extrapolations provided by the WSA model. The fourth and last PFSS model we employed in this study is the Predictive Science Inc. (PSI)-PFSS. Being built using the finite difference scheme, which is a different numerical scheme to that used in WSA and EUHFORIA, the PSI-PFSS is considered here in order to assess the extent to which the numerical set-up (for example the spherical harmonics solver versus the finite difference scheme) can affect the simulation output. Lastly, for validating the output of the PFSS-based models we consider an MHD model (hereafter referred to as PSI--MHD) that compares well to white light imaging observations of the corona \citep{mikicetal2018}. More details on all these models can be found in Appendix \ref{ap1:models}.

\subsection{Initiation global magnetic maps}
\label{sec:magnetograms}
To initialize the different coronal models, we employed solar magnetic field measurements by the Helioseismic and Magnetic Imager (HMI) \citep{Schou2012} on board the Solar Dynamics Observatory (SDO). A set of these observations covering a period of a solar rotation were processed by the photospheric flux transport model ADAPT  \citep[Air Force Data Assimilative Photospheric flux Transport --][]{Arge10, arge13, Hickmann2015, barnes2023} producing two types of line-of-sight HMI ADAPT global photospheric magnetic maps, the first with active regions (ARs) added retrospectively and the second without \citep{arge13}. These maps were generated for three consecutive date-times, as the CH we study, and which was investigated also in Paper I, was crossing the central meridional zone of the solar disk as seen from Earth's vantage point. These dates are 2010--09--18 18:00~UT, 2010--09--19 18:00~UT, and 2010--09--20 18:00~UT. For each date, the ADAPT model generates an ensemble of 12 realizations per magnetogram type (with or without ARs added retrospectively). The 12 realizations are driven by different supergranulation patterns, where the relative differences are greatest in the polar regions (i.e., above 70 degrees latitude) and the farside, especially just before the east limb data assimilation boundary. As our interest focuses on the open and closed field topology at the photosphere, both for the whole solar surface and the specific CH studied in Paper I, we were interested in an intercomparison of modeled topologies by all 12 realizations per date and per magnetogram type. In section \ref{realisation_comparison} we provide a visual and quantitative assessment to determine the extent to which the generated topologies differ from each other. For the majority, though, of our later analysis, we initialize all the models using the most optimal of the 12 realizations for each date and magnetogram type, determined based on generating accurate near-Earth solar wind conditions.

\subsection{Metrics}\label{sec:metrics}

All model simulations generate maps of the open and closed field topology at the solar surface. The maps are given in a two-dimensional grid with a resolution of 1$^{\circ}$ per pixel in longitude and latitude. Therefore, assessing their similarity and differences is easily achieved in a pixel-to-pixel comparison. In our analysis, we present difference maps that enable visual assessment of how each simulation output map compares to one another. For the CH of interest and its surrounding area, we also stack the open and closed field maps producing composite images that contribute to understanding how many of the model outputs compared agree on a specific region. To accompany these visualizations, we also provide a quantitative analysis based on a series of statistical metrics that indicate how well the modeled open and closed field areas compare to each other and to the observed EUV CH. The metrics we employ are the Jaccard similarity coefficient, marked with the symbol $J$, the overlap coefficient, marked with the symbol $O$, and the Coverage parameter introduced in \citet{asvestari19} and marked with the symbol $cov$. All metrics take values between 0 (disagree completely) and 1 (agree completely). We consider the four different comparisons given below:

\begin{itemize}
    \item Comparison of two maps with regards to their agreement for both the open and closed topology over the entire solar surface (Carrington map), which we refer to as global comparison ($J_g$, $O_g$).
    \item Comparison of two sub-maps of the area of and around the CH of interest with regards to their agreement for both the open and closed topology ($J_{g,sub}$, $O_{g,sub}$).
    \item Comparison of two sub-maps of the area of and around the CH of interest with regards to their agreement over only the open field areas ($J_{o,sub}$, $O_{o,sub}$).
    \item Comparison of the observed EUV CH area with the modeled open field area associated with that CH ($J_{CH}$, $O_{CH}$, $cov$).
\end{itemize}

For our model-observation comparisons, we use the CH boundary extracted with CATCH \citep[Collection of Analysis Tools for Coronal Holes ][]{heinemann19} and as described in Paper I. Because the closed field areas compose the largest part of the solar surface map, they bias the calculations of $J_{g}$, $O_{g}$, $J_{g,sub}$, and $O_{g,sub}$. Contrarily, when performing our analysis, it was clear that the differences between two maps were more apparent at the open field areas. Therefore, for the CH of interest we also compare the models to each other by taking into account only the open field area pixels in each sub-map as explained in the third bullet point above. The definitions and the exact formulae used for each coefficient are given in Appendix \ref{ap2:metrics}.

\subsection{The CH in brief words}\label{sec:obs_CH}

\begin{figure*}
    \centering
    \includegraphics[width=0.85\textwidth]{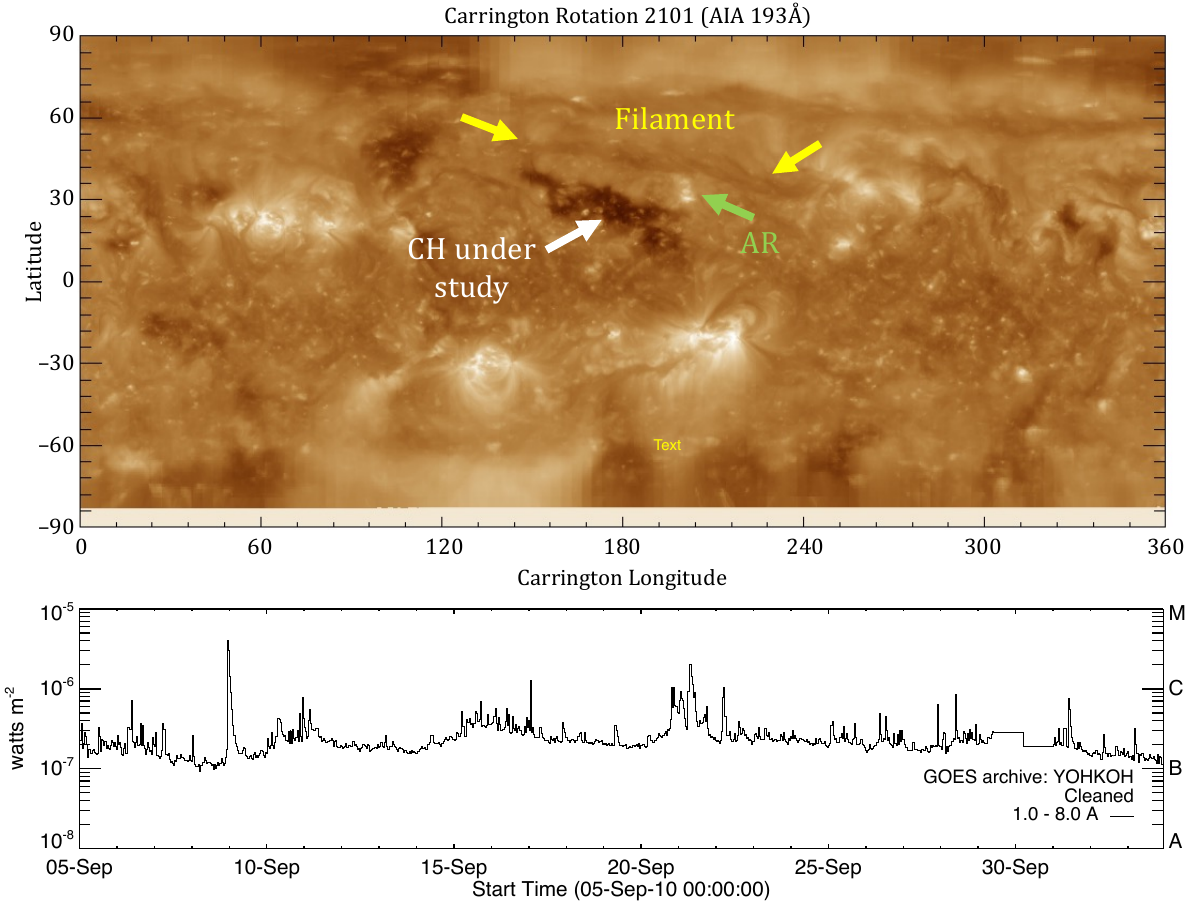}
    \caption{Top: EUV synoptic map for CR2101 using AIA 193\AA~image data giving Carrington longitude and latitude in degrees \citep[maps by][]{karna14}. The CH under study (white arrow) is surrounded by several structures. A filament channel (marked by yellow arrows) and some evolving active region (green arrow). Bottom: 5-min averaged GOES SXR flux in the 1--8\AA~channel over the time range covered by CR~2101.}
    \label{fig:CHbrief}
\end{figure*}

Figure~\ref{fig:CHbrief} shows an EUV synoptic map covering Carrington Rotation (CR) 2101, which corresponds to the period considered for modeling. The CH under study is surrounded by a filament channel, indicated by yellow arrows, and a small AR at the south-west corner of the CH, indicated by a green arrow. Several other ARs are present with enhanced EUV emissivity indicating that some had short episodes of flare activity. The bottom panel of Figure~\ref{fig:CHbrief} gives the GOES SXR flux during CR 2101. It shows some flux enhancement towards C-class flares. Several poor and very poor CME events are reported at CDAW (NASA CME catalogue~{\url{https://cdaw.gsfc.nasa.gov/CME_list/}}) for that time range, and three partial halos (Earth view). The automatized HEK (Heliophysics Events Knowledgebase~{\url{https://www.lmsal.com/hek/index.html}}) reports low energetic flare events but no flux emerging region. In general, we can conclude that the activity during CR 2101 is on a moderate level, and thus no drastic changes are anticipated to occur on the CH shape and topology over the 3--day interval we study.

\section{Results}
\label{sec_results}

\subsection{Assessing the impact of the input global magnetic maps}
\label{magnetograms}

The choice of the global magnetic map used to initiate the model simulations can lead to significantly different output when it comes to the open and closed magnetic field topology and field strength. This will subsequently affect the computed plasma parameters (density, speed, and temperature) but investigating this falls out of the scope of this paper. However, we are interested in quantifying to what extend employing a different global magnetic map alters the resulting open and closed field topology. Although other studies have investigated the impact of global magnetic maps produced by different instruments/techniques \citep[see for example][]{riley2014, lietal2021}, we focus on assessing (1) the difference/similarities between the PFSS modeled open and closed field topology maps initiated by the 12 different HMI ADAPT realizations and (2) the effect of the ARs when they are added retrospectively in the HMI ADAPT global magnetic maps.

\subsubsection{Assessing the results initiated by different HMI ADAPT realizations}
\label{realisation_comparison}

\begin{figure*}[h!]
    \centering
    \includegraphics[width = 0.75\textwidth]{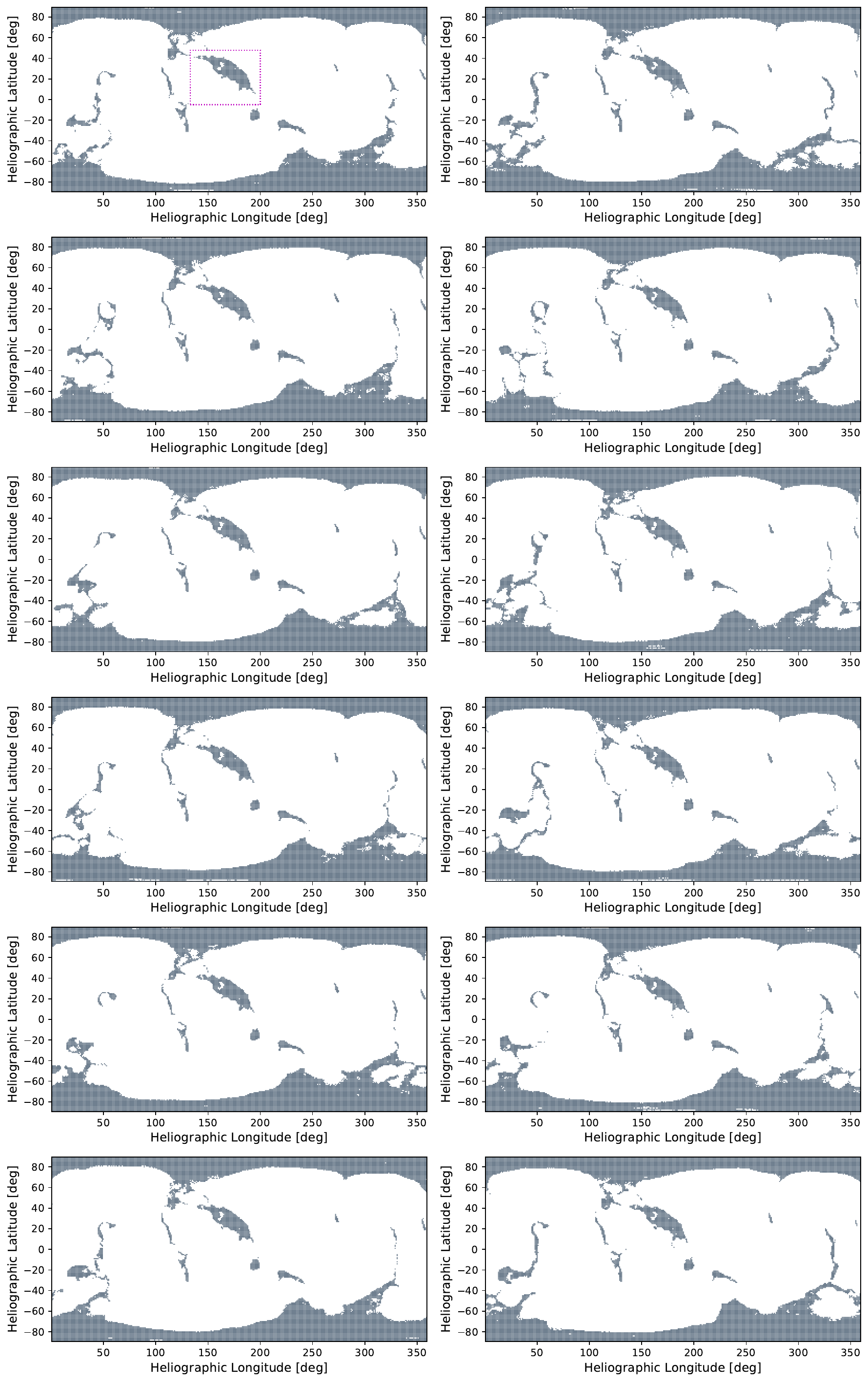}
    \caption{Open and closed field topology maps for 2010-09-19 produced by the WSA model that was initiated with each of the HMI ADAPT 12 realizations with ARs added retrospectively. The realizations are presented row-wise from realization 1 at the top left to realization 12 at the bottom right panel. The magenta dotted rectangle in the top left panel indicates the area that corresponds to the CH we study here and in Paper I.}
    \label{fig:20100919_with_realisation_maps}
\end{figure*}

For each of the three dates, we study 12 HMI ADAPT map realizations which are utilizing different supergranulation patterns \citep[see for example,][]{Hickmann2015, barnes2023}. Although for the bulk of our study we employed the most optimal realization provided by the ADAPT model, where the optimization is based on their assessment of how accurately the WSA simulation output reproduces in situ solar wind measurements at L1, we quantify how the PFSS modeled topology will differ if any of the other realizations was selected. We obtained the open and closed field topology using the WSA model initiated using each of the 12 modeled realizations. This was done both for global magnetic maps where ARs were added retrospectively and maps without ARs added. Figure \ref{fig:20100919_with_realisation_maps} shows all the simulation output maps of the open and closed field topology generated by the WSA model for the 2010-09-19 initiated with the HMI ADAPT global magnetic maps with ARs added retrospectively. In the figure the realizations are presented row-wise from realization 1 at the top left to realization 12 at the bottom right panel. The same figures for the rest of the dates and for the case of using the map without ARs added are given as supplementary material. The magenta dotted rectangle indicates the open field area that corresponds to the CH that is the focus of this study. One can see that there are differences between the maps, but they are not very strong, in particular around the specific CH of interest. 

To understand how significant these differences are, we make map-to-map comparisons, both for the entire Carrington map and for a sub-map that focuses on the specific CH we study and the area surrounding it (the area enclosed by the magenta rectangle in Figure \ref{fig:20100919_with_realisation_maps}). For a given date and a given type of global magnetic map (with or without ARs added retrospectively) we computed the Jaccard similarity and the Overlap coefficients comparing all 12 realizations with each other. This comparison produced 76 values for each coefficient per date and type of global magnetic map considered. The mean value and its standard error obtained are given in Table \ref{tab:mean_metrics_realisations}. The resulted coefficients $J_{g}$, $O_{g}$ scored above 0.9 suggesting that all realizations produced maps that agree well with each other. When focusing on the specific CH area studied here, the $J_{o,sub}$, as expected, drops compared to the $J_{g,sub}$, $O_{g,sub}$, and $O_{o,sub}$, but still the values remain above 0.9. From this we can conclude that using any of the 12 realizations for a given date and type of magnetogram leads to highly comparable open and closed field topologies.

\begin{figure*}[h!]
\centering
    \begin{minipage}{0.99\textwidth}
    \centering
        \includegraphics[width = 0.75\textwidth]{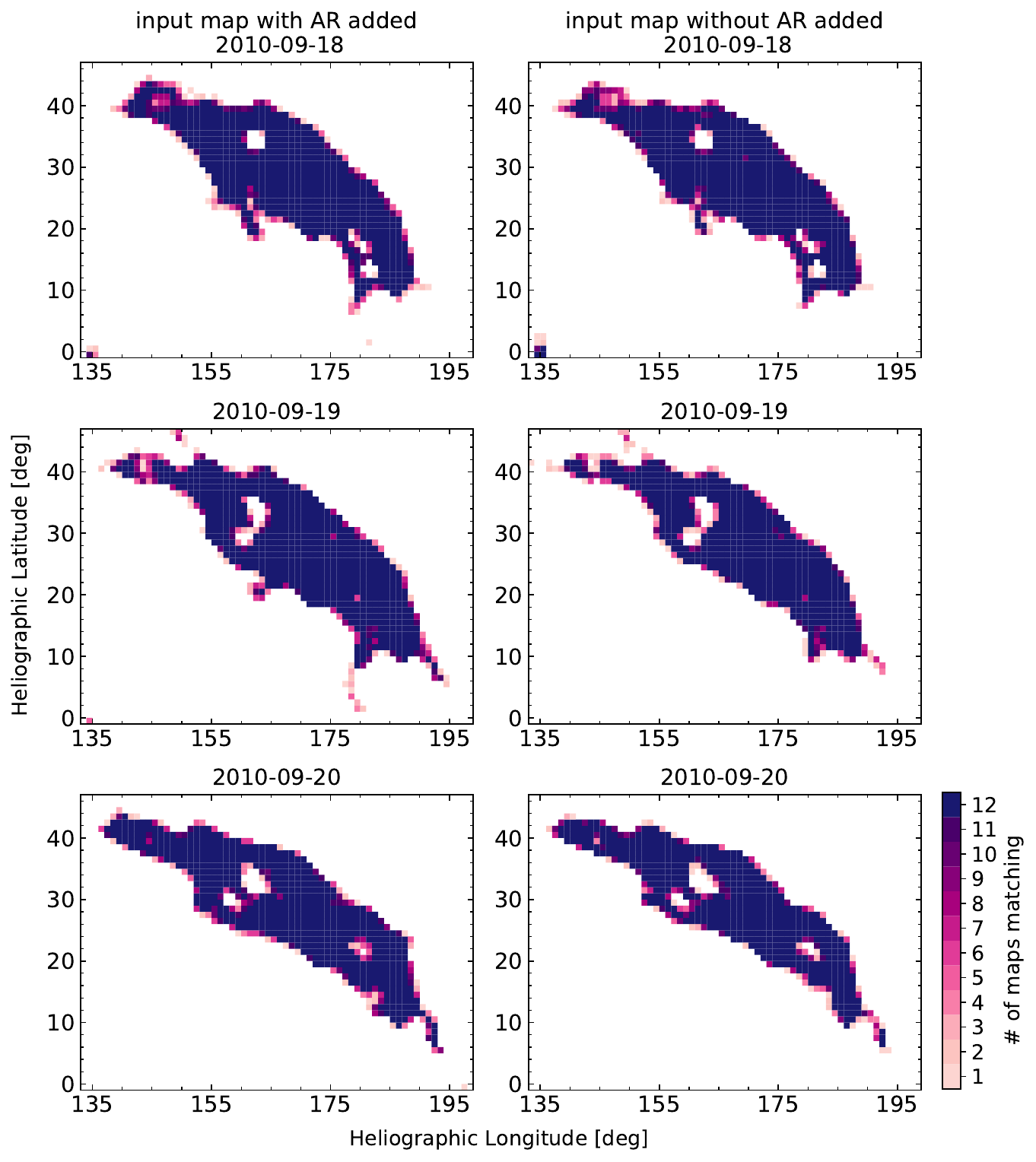}
    \end{minipage}
    \caption{Stacked maps of the  WSA model output of opened and closed field topology generated by the different HMI ADAPT realization at a given date (row) using maps with ARs (left column) and without ARs (right column) added retrospectively. The colormap indicates the number of modeled realization agreeing at a given pixel on the grid, ranging from 1 (light pink) to 12 maps (dark purple).}
    \label{fig:20100919_with_without_realisations_stacked}
\end{figure*}

To further visualize how well the maps agree or disagree at the CH area of interest, stacked maps of the WSA model output produced based on the 12 realizations per day and magnetogram type are given in Figure \ref{fig:20100919_with_without_realisations_stacked}. The left column shows the stacked maps generated by initializing the WSA model with the global magnetic maps with ARs added retrospectively, while the right column with those without ARs added. Each row shows the results for a different date. The color map, varying from light pink to dark purple, shows, increasingly from 1 to 12, the number of maps agreeing at a given pixel. Throughout all figure panels it is clear that when ARs are added the resulting WSA maps differ comparing to those from runs without ARs added (left versus right panels of figure -- we address this in detail in the following section). However, when looking at each individual panel one can see that not much variation is present, suggesting that the output from all 12 realizations agree for the majority of the CH region. This supports our conclusion that even though for the remaining of the study we used what was considered as the most optimal realization per date and magnetogram type, using any of the other realizations would have produced highly comparable results. At this point it is important to mention that modeled open and closed field topology generated using magnetograms produced by different instruments/techniques may lead to a larger variation among the results. This has been extensively studied in, for example, \citet{riley2014, lietal2021}. In this section our focus was solely on understanding whether variation is generated among the runs based on the different realizations of the ADAPT magnetic maps we considered.

\begin{table*}
\centering
\caption{Metrics for comparing the model outputs initiated by the 12 realizations}\label{tab:mean_metrics_realisations}
\begin{tabular}{l  c  c  c  c  c  c }
\hline

\hline
\multicolumn{7}{c}{Input global magnetic map with AR added retrospectively }\\
\hline
date & $J_g$ & $O_g$ & $J_{g,sub}$ & $O_{g,sub}$ & $J_{o,sub}$ & $O_{o,sub}$\\ \hline
2010-09-18 & 0.91$\pm0.001$ & 0.95$\pm0.001$ & 0.96$\pm0.001$ & 0.98$\pm0.001$ & 0.92$\pm0.002$ & 0.97$\pm0.001$ \\
2010-09-19 & 0.91$\pm0.001$ & 0.95$\pm0.0004$ & 0.96$\pm0.001$ & 0.98$\pm0.0004$ & 0.92$\pm0.002$ & 0.97$\pm0.001$ \\
2010-09-20 & 0.92$\pm0.001$  & 0.96$\pm0.001$  & 0.96$\pm0.001$  & 0.98$\pm0.0004$  & 0.92$\pm0.002$  & 0.98$\pm0.001$  \\
\hline

\hline
\multicolumn{7}{c}{Input global magnetic map without AR added retrospectively }\\
\hline
date & $J_g$ & $O_g$ & $J_{g,sub}$ & $O_{g,sub}$ & $J_{o,sub}$ & $O_{o,sub}$\\ \hline
2010-09-18 & 0.91$\pm0.0008$ & 0.96$\pm0.0004$ & 0.96$\pm0.001$ & 0.98$\pm0.0004$ & 0.91$\pm0.002$ & 0.97$\pm0.001$ \\
2010-09-19 & 0.92$\pm0.001$ & 0.96$\pm0.0004$ & 0.96$\pm0.001$ & 0.98$\pm0.001$ & 0.91$\pm0.0003$ & 0.97$\pm0.001$ \\
2010-09-20 & 0.92$\pm0.001$ & 0.96$\pm0.0004$ & 0.97$\pm0.001$ & 0.98$\pm0.0003$ & 0.92$\pm0.002$ & 0.97$\pm0.001$ \\ \hline

\hline
\end{tabular}
\tablecomments{Mean value and error calculated for the Jaccard similarity and the Overlap coefficients described in Appendix \ref{ap2:metrics} over the whole map and the sub-map area for each day and global magnetic map type based on map-to-map comparisons of the WSA output initiated by the 12 realizations. This assesses how well the model outputs agree with each other on the reconstructed open and closed field areas.}
\end{table*}

\subsubsection{Assessing the results generated by the two global magnetic map types}
\label{magnetogram_type_comparison}

In this section, we compare the produced magnetic field topology by a given model implementation for a given date when the model is initiated by the two types of HMI ADAPT maps, with and without ARs added. The analysis was applied to all four PFSS based models. For a given date and a given model we compared the topology output when using the two aforementioned global magnetic maps (model intra-comparison). Using equations \ref{eq:jaccard_global} and \ref{overlap_global}, respectively, we calculated the Jaccard similarity and the Overlap coefficients. The results are given in rows 3--6 and columns 2--4 and 5--7 of Table \ref{tab:with_without}, respectively. By combining all model intra-comparisons for a given date we estimated the mean value and its error, which are given in row 7, columns 2--4 and 5--7 of the same table. Although, it is expected that ARs can influence the output of global magnetic field models this is only moderately reflected in our simulation outputs for the considered global magnetic maps. The metrics for each model intra-comparison but also the mean values for a given date indicate that there are only small differences between the topologies generated by the two maps. The remaining table rows and columns show the metrics calculated for the sub-map regions, which also support the above argument.

\begin{figure*}[h!]
\centering
\centering
    \begin{minipage}{0.99\textwidth}
    \centering
    \includegraphics[width = 0.99\textwidth]{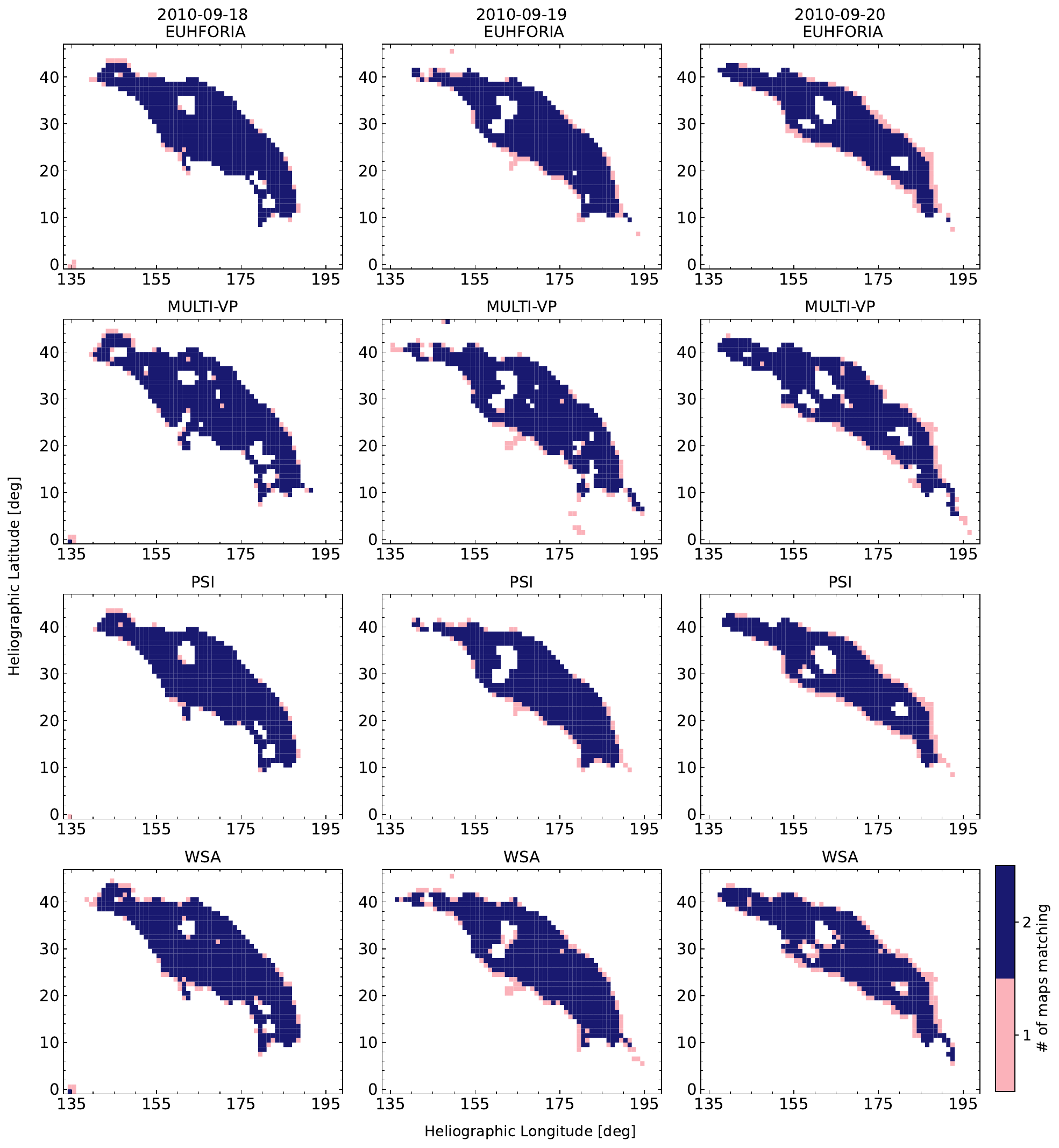}
    \end{minipage}
    \caption{Stacked model output of open and closed field topology generated by EUHFORIA (first row), Multi-VP (second row), PSI-PFSS (third row), and WSA (fourth row) for a given date per column by the two types of HMI ADAPT maps considered in the study, with and without ARs added retrospectively. Light pink color means the maps do not agree, while dark purple that they do agree for the given pixel.}
    \label{fig:with_without_intercomparison}
\end{figure*}

Figure \ref{fig:with_without_intercomparison} shows stacked images of the sub-maps produced for a given date by overlaying the model output based on the two input global magnetic maps. Each column shows the results for a different day and each row for a different model. Light pink color means that only one map produced open field topology at a given pixel (maps do not  have overlapping open field pixels), while dark purple means that both maps match at a given pixel. Subtle differences between the two areas are visible predominately at the boundary of the CH. The maps produced for 2010-09-19 and 2010-09-20 are showing the largest difference between the output of a given model based on the two global magnetic maps compared to the output from 2010-09-18.

\begin{table*}
\centering
\caption{Intra-comparison of the same model output initiated by the two different type global magnetic maps.}\label{tab:with_without}
\centering
\begin{tabular}{ l  c  c  c || c  c  c }
\hline

\hline
 & \multicolumn{3}{c||}{$J_g$} & \multicolumn{3}{c}{$O_g$}\\ \cline{2-7}
 & 2010-09-18 & 2010-09-19 & 2010-09-20 & 2010-09-18 & 2010-09-19 & 2010-09-20\\ \hline
EUHFORIA & 0.89 & 0.90 & 0.89 & 0.94 & 0.95 & 0.94 \\
MULTI-VP & 0.88 & 0.88 & 0.88 & 0.93 & 0.94 & 0.94 \\
PSI-PFSS & 0.89 & 0.90 & 0.90 & 0.94 & 0.95 & 0.95 \\
WSA & 0.88 & 0.89 & 0.89 & 0.94 & 0.94 & 0.94 \\
Mean value & 0.88$\pm0.004$ & 0.89$\pm0.004$ & 0.89$\pm0.003$ & 0.94$\pm0.002$ & 0.94$\pm0.002$ & 0.94$\pm0.002$ \\
\hline

\hline
 & \multicolumn{3}{c||}{$J_{g,sub}$} & \multicolumn{3}{c}{$O_{g,sub}$}\\ \cline{2-7}
 & 2010-09-18 & 2010-09-19 & 2010-09-20 & 2010-09-18 & 2010-09-19 & 2010-09-20\\
EUHFORIA & 0.98 & 0.96 & 0.95 & 0.99 & 0.98 & 0.97 \\
MULTI-VP & 0.97 & 0.96 & 0.96 & 0.99 & 0.98 & 0.98 \\
PSI-PFSS & 0.98 & 0.97 & 0.95 & 0.99 & 0.98 & 0.97 \\
WSA & 0.97 & 0.95 & 0.94 & 0.98 & 0.97 & 0.97 \\
Mean value & 0.98$\pm0.004$ & 0.96$\pm0.005$  & 0.95$\pm0.004$ & 0.99$\pm0.002$ & 0.98$\pm0.002$ & 0.97$\pm0.002$ \\
\hline

\hline
 & \multicolumn{3}{c||}{$J_{o,sub}$} & \multicolumn{3}{c}{$O_{o,sub}$}\\ \cline{2-7}
 & 2010-09-18 & 2010-09-19 & 2010-09-20 & 2010-09-18 & 2010-09-19 & 2010-09-20\\
EUHFORIA & 0.94 & 0.89 & 0.84 & 0.99 & 0.98 & 1.00 \\
MULTI-VP & 0.93 & 0.88 & 0.87 & 0.97 & 0.97 & 0.99 \\
PSI-PFSS & 0.95 & 0.90 & 0.83 & 0.99 & 0.98 & 1.00 \\
WSA & 0.91 & 0.86 & 0.82 & 0.97 & 0.96 & 0.98 \\
Mean value & 0.93$\pm0.007$ & 0.88$\pm0.008$ & 0.84$\pm0.011$ & 0.98$\pm0.006$ & 0.97$\pm0.004$ & 0.99$\pm0.005$ \\ \hline

\hline
\end{tabular}
\tablecomments{The global magnetic maps considered were the HMI-ADAPT with and without ARs added retrospectively.}
\end{table*}

\subsection{Assessing the impact of the simulation set up}

As discussed in the introduction, coronal models are sensitive to the simulation set-up and the numerical schemes employed. In this section, we focus on the effect of global magnetic map smoothing and of the selection of the source surface height. Furthermore, by comparing all models to each other we assess the overall effect of the different numerical solution schemes implemented.

\subsubsection{Assessing the results based on smoothing the global magnetic map }
\label{smoothing_comparison}

\begin{figure*}[t!]
\centering
\centering
    \begin{minipage}{0.99\textwidth}
        \includegraphics[width = 0.99\textwidth]{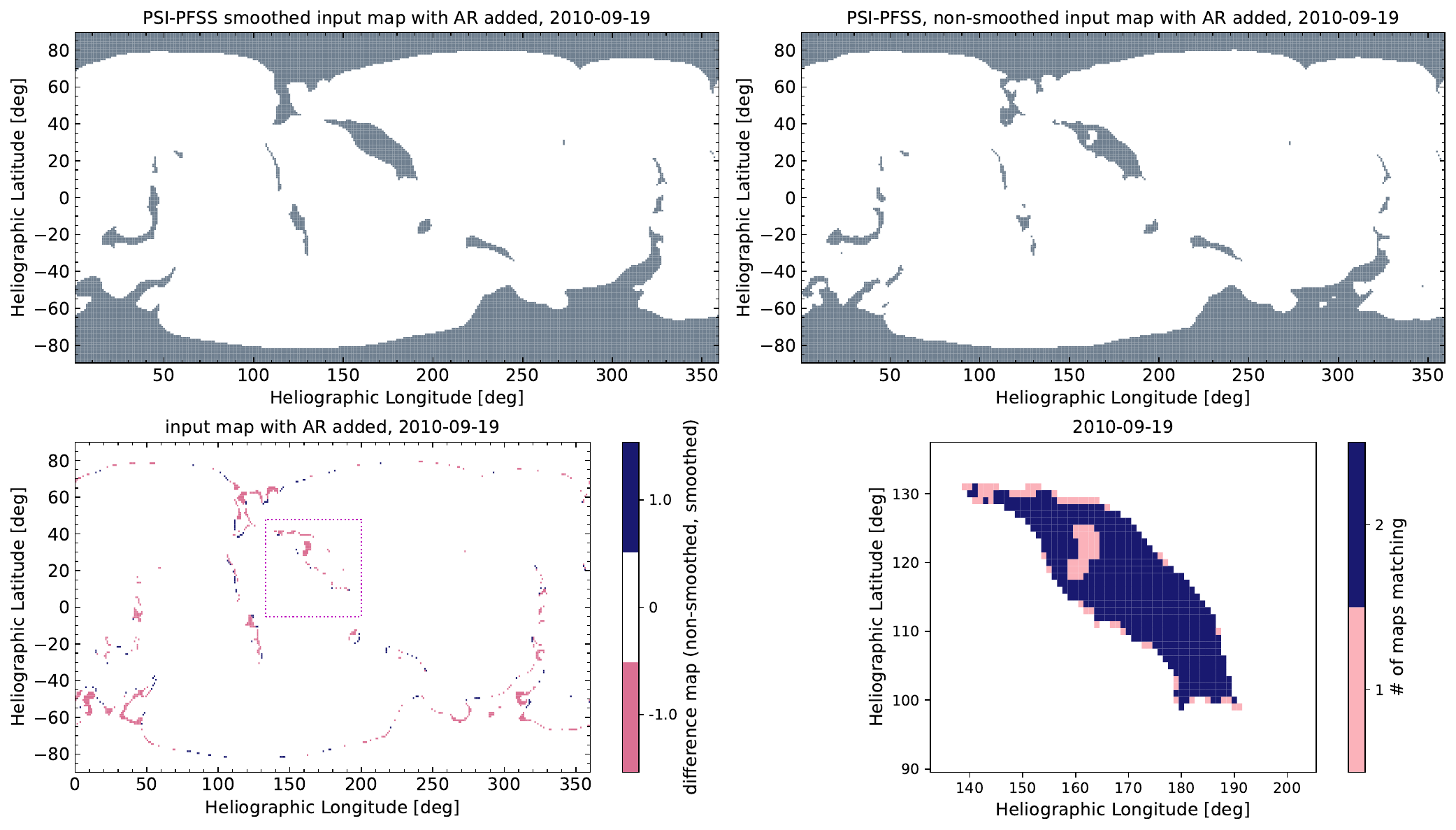}
    \end{minipage}
    \caption{Top row: Modeled  open (gray) and closed (white) magnetic field topology for the 2010-09-19 using the PSI-PFSS initiated with a smoothed (left) and non-smoothed (right) HMI ADAPT map for the 2010-09-19. Bottom row: Difference Carrington map (left) and stacked model output (right) of open and closed field topology of the CH region of the two maps shown above. The dashed magenta rectangle on the left panel shows the area for which the stacked map is generated.
    }
    \label{fig:PSI_PFSS_smoothed}
\end{figure*}

Usually global magnetic maps are smoothed before they are input into a given model. This can affect the modeled magnetic field topology. In this section, we compare two model outputs produced by the PSI-PFSS model, one produced by smoothing the input map (Figure~\ref{fig:PSI_PFSS_smoothed} -- top row, left panel) and one without smoothing (Figure~\ref{fig:PSI_PFSS_smoothed} -- top row, right panel). The global magnetic map utilized is the HMI ADAPT with ARs added retrospectively for the date of 2010-09-19. The two maps appear to be visibly similar, but there are clear differences near regions of open field. More precisely, fine structures located within open field areas or at their boundaries disappear when the input global magnetic map is smoothed. Small areas of open field are enlarged after the map is smoothed, and some neighboring ones merge. Despite these areas of discrepancy between the two maps, the global Jaccard similarity and Overlap coefficients are scoring high when we make map-to-map comparisons of the entire map. Namely $J_g = 0.97$ and $O_g = 0.99$. The left panel of the bottom row of Figure~\ref{fig:PSI_PFSS_smoothed} shows a difference map between the two model outputs with the dashed magenta rectangle marking the area for which the stacked map on the right panel of the same row was produced. Both panels highlight the loss of the open field area fine structures when smoothing is applied to the global magnetic map. For the sub-map the metrics are estimated to be $J_{g,sub} = 0.95$, $O_{g,sub} = 0.98$, $J_{o,sub} = 0.86$, and $O_{o,sub} = 0.99$. Despite the differences between the two maps, the metrics score relatively high, which suggests that the differences are localized and relatively small. For the sub-maps when focusing the comparisons only on the open field areas the $J_{o,sub}$ takes a smaller value, strongly showing how biased the other metrics are to the presence of closed field areas, which occupy the majority of the maps and sub-maps.

\subsubsection{Assessing the results generated by different source surface heights}
\label{height_comparison}

\begin{figure*}[h!]
\centering
    \begin{minipage}{0.99\textwidth}
        \includegraphics[width = 0.99\textwidth]{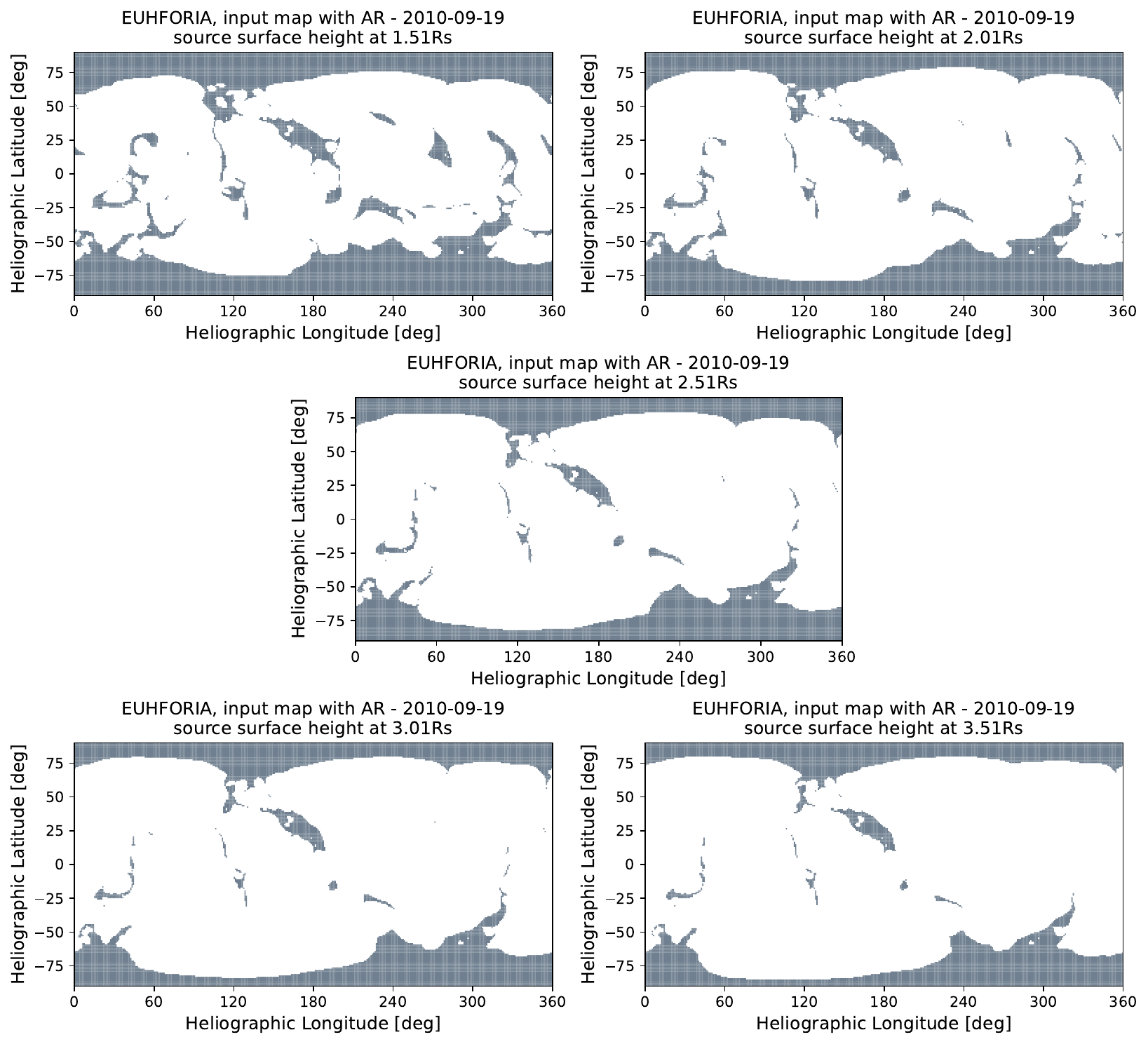}
    \end{minipage}
    \caption{Open and closed field topology generated by EUHFORIA for the 2010--09--19 when the model was initiated with different source surface and inner SCS model boundary heights each time (top-left: [1.49, 1.51]$R_s$, top-right: [1.99, 2.01]$R_s$, middle: [2.49, 2.51]$R_s$, bottom-left: [2.99, 3.01]$R_s$, and bottom-right: [3.49, 3.51]$R_s$). The input map used here is the HMI ADAPT with ARs added retrospectively.}
    \label{fig:dif_source_surface_heights_maps}
\end{figure*}

When modeling the open and closed field topology with a PFSS-SCS model the selection of the height of the outer boundary of the PFSS model (the source surface height) and subsequently the height of the inner boundary of the SCS model can significantly affect the size of the modeled open and closed field areas. More precisely, lowering the boundary heights will result in increasing the size of areas of open field, while rising the boundary heights will lead to reducing it. In this subsection, we quantify the similarity of the generated open and closed field maps when the model is initiated using different boundary heights. For this, we used the EUHFORIA coronal model and generated simulations using the HMI ADAPT map with ARs added retrospectively for 2010-09-19. The source surface and the inner SCS boundary height pairs considered in this analysis are [1.49, 1.51]$R_s$, [1.99, 2.01]$R_s$, [2.49, 2.51]$R_s$, [2.99, 3.01]$R_s$, and [3.49, 3.51]$R_s$, where the third pair listed is the one that had been used throughout this study by all the PFSS models and is hereafter referred to as the nominal values. The EUHFORIA simulation output based on these 5 boundary height pairs is given in Figure \ref{fig:dif_source_surface_heights_maps}, where the top row shows the model output for boundary heights below the nominal values, the middle row gives the modeled map for the nominal values of boundary heights, while the bottom row gives the output maps for boundary heights above the nominal values. As expected our results are in agreement with \citet{asvestari19} and \citet{caplanetal2021}. Namely, when the boundaries of the simulation domains were lowered the open field areas, indicated in gray in the figure, grew significantly. In addition, open field areas appeared in locations where they were not present before. Raising the boundary heights led accordingly to a decrease of the open field areas, but the decrease is not as strong as the increase in the cases of lowered boundary heights. 

\begin{figure*}[h!]
\centering
\centering
    \begin{minipage}{0.99\textwidth}
        \includegraphics[width = 0.99\textwidth]{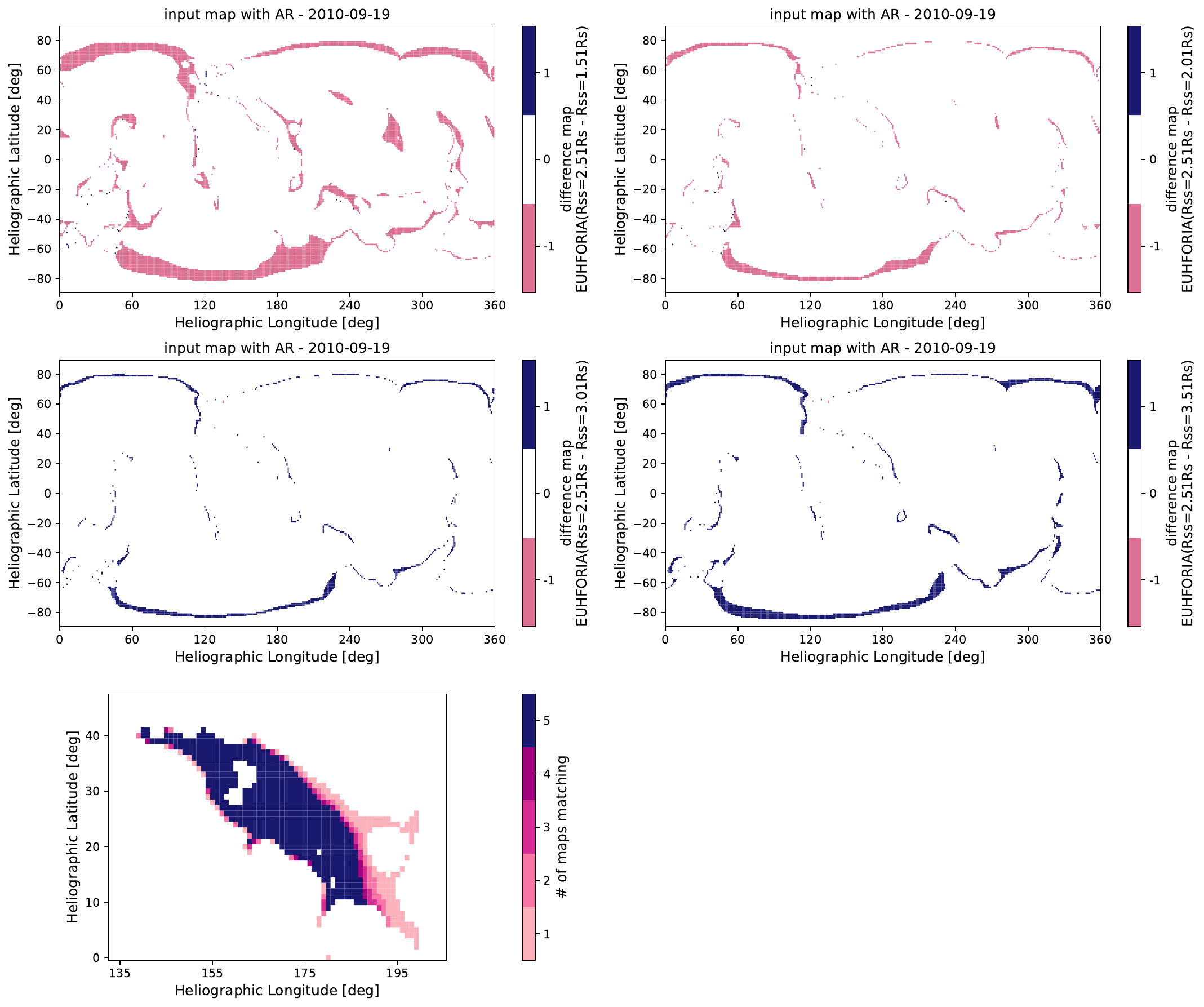}
    \end{minipage}
    \caption{Top and middle row: Difference maps of EUHFORIA simulation output initiated with the nominal values for the boundary heights [2.49, 2.51]$R_s$ and the simulation output initiated with the other three pairs (top-left: [1.49, 1.51]$R_s$, top-right: [1.99, 2.01]$R_s$, middle-left: [2.99, 3.01]$R_s$, and middle-right: [3.49, 3.51]$R_s$). Bottom row: Stacked model output of opened and closed field topology generated by EUHFORIA when the model was initiated the different pair of boundary heights. The input global magnetic map used here is the HMI ADAPT with ARs added retrospectively for the 2010-09-19. The colourmap indicates the number of modeled outputs agreeing at a given pixel on the grid.}
    \label{fig:dif_source_surface_heights_intercomparison}
\end{figure*}

To visualize this better, we produce the difference maps between the modeled topology for the nominal boundary heights, [2.49, 2.51]$R_s$, and each of the other maps. The difference maps are presented in rows 1--2 of Figure \ref{fig:dif_source_surface_heights_intercomparison}. One can see that by lowering the boundary heights significantly below the nominal values, the areas of disagreement between the two maps are larger than when the boundary heights were increased above the nominal values. The polar open field regions show the most dramatic change in the area when changing the boundary heights. One can see from the same difference maps that the simulation output based on the lowest selected boundary heights show the strongest difference to the nominal values and subsequently to heights above them. This is reflected in the calculated global Jaccard similarity and the Overlap coefficients over the whole map comparisons and the coefficients for the sub-maps given in Table~\ref{tab:topology_comparisons_dif_heights}. The metrics shown in this table are calculated for map-to-map comparisons between the map produced with the nominal boundary heights and each of the other maps. As can be seen, all the metrics score above 0.9 suggesting a good map agreement, apart from the comparison with the output initialized with the lowest boundary heights as suggested already from the visual comparisons.

A stacked image of all simulated maps from the runs with different boundary heights is given at the bottom row of Figure~\ref{fig:dif_source_surface_heights_intercomparison}. From this image and the four panels in the images above, one can conclude that decreasing the boundary heights shows an increase in the open area associated with the CH we study at its southern edge and towards the solar west. The region towards which the open field area grows lies below an active region. We note that the southern hemisphere is characterized by ARs 11106 and 11108 at this time. A key output is that the selection of the two boundary heights has the greatest impact on the modeled topologies, in contrast to the comparisons made in the previous sections.

\begin{table*}
{\centering
\caption{Similarity metrics for the EUHFORIA simulation output based on different boundary heights}\label{tab:topology_comparisons_dif_heights}
\centering
\begin{tabular}{ l  c  c  c  c  c  c }
\hline

\hline
    Source surface heights & \hspace*{6mm} $J_g$ \hspace*{6mm} & $O_g$ & $J_{g,sub}$ & $O_{g,sub}$  & $J_{o,sub}$ & $O_{o,sub}$\\ \hline
    2.51$R_s$ vs. 1.51$R_s$ & 0.82 & 0.90 & 0.85 & 0.92 & 0.76 & 1.0 \\
    2.51$R_s$ vs. 2.01$R_s$ & 0.93 & 0.96 & 0.91 & 0.95 & 0.93 & 1.0 \\
    2.51$R_s$ vs. 3.01$R_s$ & 0.95 & 0.97 & 0.92 & 0.96 & 0.95 & 1.0 \\
    2.51$R_s$ vs. 3.51$R_s$ & 0.91 & 0.96 & 0.91 & 0.95 & 0.92 & 1.0\\ \hline

    \hline
    \end{tabular}}
\end{table*}

\subsubsection{Assessing the result from the different PFSS models}
\label{PFSS_intercomp}

\begin{figure*}[t!]
\centering
\centering
    \begin{minipage}{0.99\textwidth}
            \includegraphics[width = 0.98\textwidth]{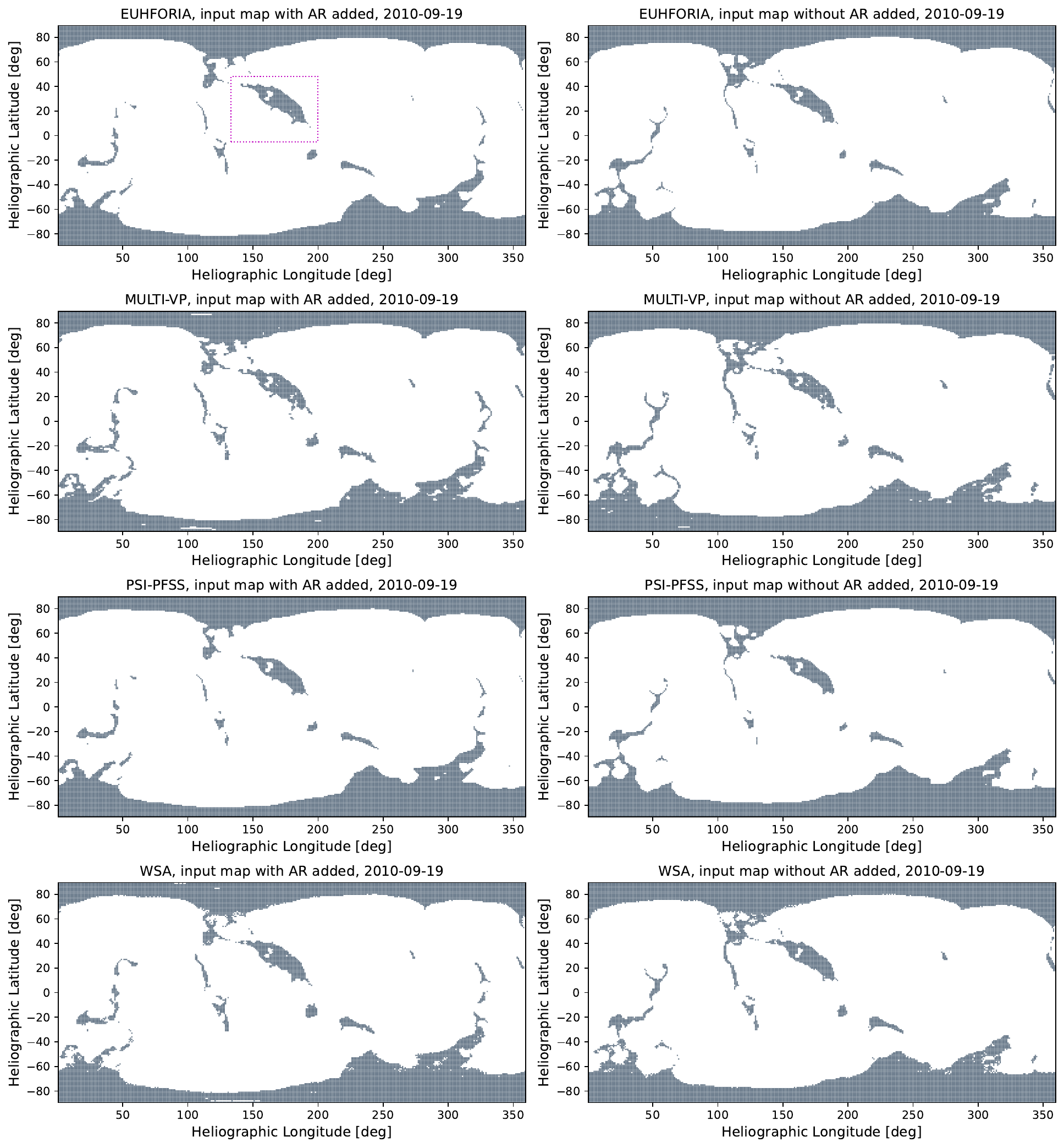}
    \end{minipage}
    \caption{Modeled open (gray) and closed (white) magnetic field topology for the 2010-09-19 using EUHFORIA (top row), MULTI-VP (second row) , PSI-PFSS (third row), and WSA (last row) models described in Appendix~\ref{ap1:models}. Each model was initiated with the optimal realizations for the HMI ADAPT map with (left column) and without (right column) AR added retrospectively.}
    \label{fig:20100919_with_maps}
\end{figure*}

\begin{figure*}[t!]
\centering
\centering
    \begin{minipage}{0.95\textwidth}
    \includegraphics[width = 0.95\textwidth]{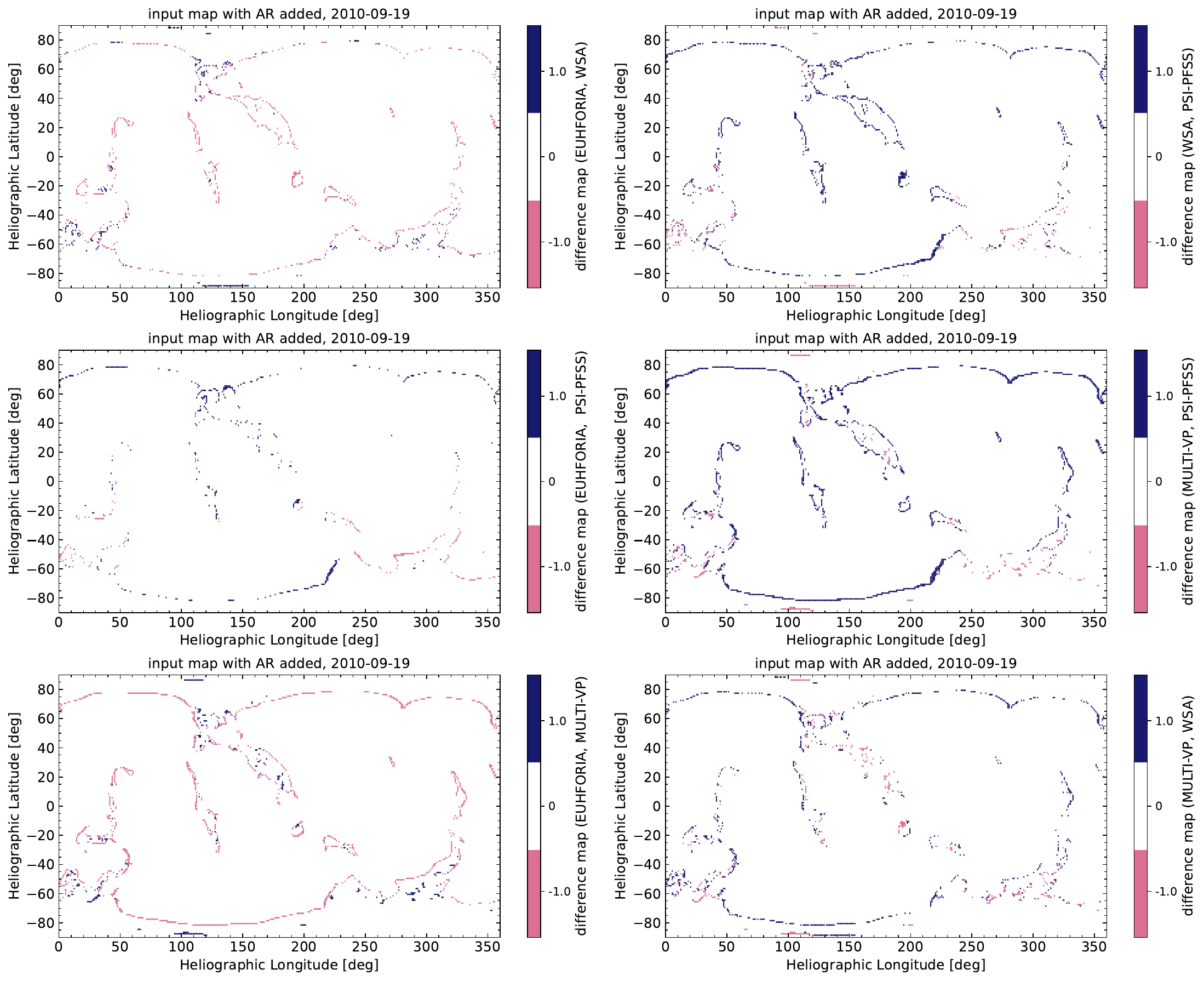}
    \end{minipage}
    \caption{Difference maps between the four PFSS models considered in this study in the case of being initiated with the HMI ADAPT maps with ARs added retrospectively for the 2010--09--19.}
    \label{fig:20100919_with_difference_maps}
\end{figure*}

\begin{figure*}[t!]
\centering
\centering
    \begin{minipage}{0.95\textwidth}
    \includegraphics[width = 0.95\textwidth]{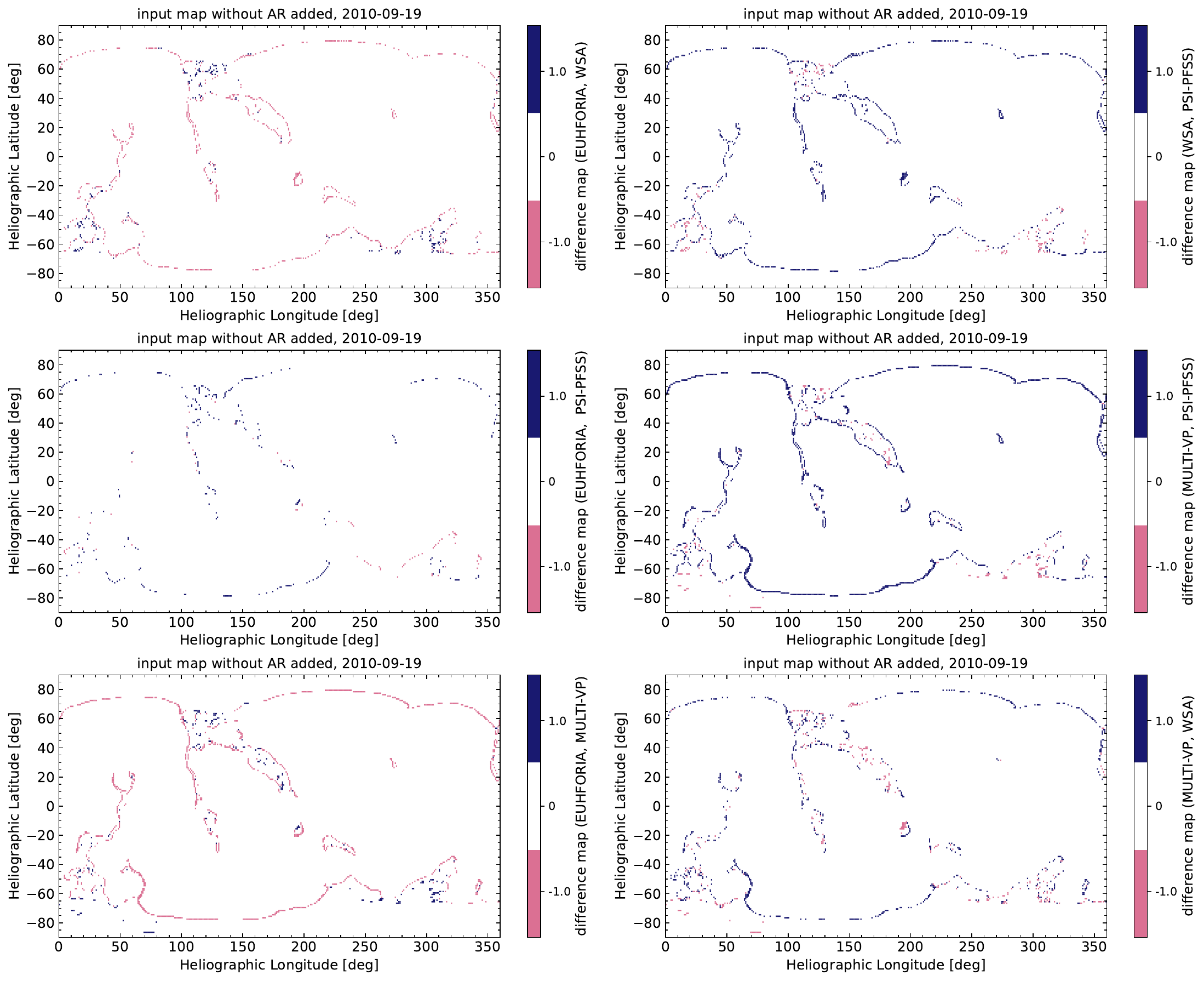}
    \end{minipage}
    \caption{Difference maps between the four PFSS models considered in this study in the case of being initiated with the HMI ADAPT maps without ARs added retrospectively for the 2010--09--19.}
    \label{fig:20100919_without_difference_maps}
\end{figure*}

To assess to what extent the numerical schemes employed by the different implementations of the models lead to different open and closed field topologies, in this section, we compare all PFSS-based models to each other. At this point it is worth reminding the reader that EUHFORIA and WSA use a spherical harmonic solver, MULTI-VP employs the WSA solution but then processes the data further \citep[for details see][]{pinto17}, and PSI-PFSS uses a finite difference scheme. Another added difference among the models is the method used for field line tracing. All Carrington maps of open and closed field topology for 2010-09-19 are presented in Figure~\ref{fig:20100919_with_maps}. The left column presents maps produced by global magnetic maps with ARs added retrospectively and the right column those produced by maps without ARs. Each row shows the result from a given model, and the models are listed alphabetically row-wise. Visual comparison of the maps does not indicate significant differences among the models. A subtle difference can be identified which is better seen in the difference maps shown in Figure \ref{fig:20100919_with_difference_maps}, for maps generated by global magnetic maps with ARs added retrospectively, and Figure \ref{fig:20100919_without_difference_maps} by maps without ARs added. These figures show the results for 2010-09-19, for the other two dates the same images are provided as supplementary material. We can see in both figures that the maps produced by MULTI-VP are consistently different from those produced by EUHFORIA and PSI models but agree better with the WSA one. This is expected considering that the solution of the latter was used to initiate it. Overall, one can distinguish thin bands at the boundaries both of open field regions at the poles and at different latitudes, which indicate that the MULTI-VP produces larger areas of open field compared to all other models. EUHFORIA and PSI outputs show the best agreement with each other, while both also compare relatively well to the WSA output.

The metrics for the map comparisons for all three dates are given in Tables~\ref{tab:open_closed_topology_comparisons} and \ref{tab:open_closed_topology_comparisons_without}. The estimated values are relatively high, above 0.9, when comparing the entire solar surface maps (columns 2 and 3 of both tables) both for open and closed fields, which is in accordance with the visual agreements of all the maps in Figures \ref{fig:20100919_with_difference_maps} and \ref{fig:20100919_without_difference_maps}. In the case of sub-map comparisons around the CH of interest the $J_{g,sub}$ and $O_{g,sub}$ metrics given in columns 4 and 5 of the two tables show a slight drop but they still remain above 0.9. When comparing only the open field areas, the metric $J_{o,sub}$ given in column 6 of the same tables drops even more, with some values being below 0.8, showing that the maps agree less with each other for that CH. The metric $O_{o,sub}$, however, scores high, namely above 0.9. Essentially, when comparing the open field areas only with this metric, one assesses if the smallest area of open field simulated by one of the two models being compared is one-to-one captured by the other model, if this condition is fully met then the metric obtains the value 1.00. However, if a portion of the smallest area is not captured by the other model then the metric drops below 1.00.

\begin{table}
\centering
\caption{Similarity metrics from mode-to-model comparisons of the open and closed field topologies simulated using the HMI ADAPT maps with ARs added retrospectively.} \label{tab:open_closed_topology_comparisons}
{\centering
\begin{tabular}{ l  c  c  c  c  c  c }
\hline

\hline
\multicolumn{7}{c}{2010--09--18} \\ \hline
 & $J_g$ & $O_g$ & $J_{g,sub}$ & $O_{g,sub}$ & $J_{o,sub}$ & $O_{o,sub}$\\ \hline
EUHFORIA vs. MULTI--VP  & 0.96 & 0.98 & 0.94 & 0.97 & 0.83 & 0.93 \\
EUHFORIA vs. PSI--PFSS  & 0.99 & 1.00 & 0.98 & 0.99 & 0.94 & 0.98 \\
EUHFORIA vs. WSA        & 0.97 & 0.99 & 0.96 & 0.98 & 0.88 & 0.99 \\
MULTI--VP vs. PSI--PFSS & 0.95 & 0.98 & 0.93 & 0.97 & 0.81 & 0.93 \\
MULTI--VP vs. WSA       & 0.97 & 0.98 & 0.96 & 0.98 & 0.88 & 0.96 \\
PSI--PFSS vs. WSA       & 0.97 & 0.98 & 0.95 & 0.97 & 0.86 & 0.99 \\ \hline

\hline
\multicolumn{7}{c}{2010--09--19} \\ \hline
 & $J_g$ & $O_g$ & $J_{g,sub}$ & $O_{g,sub}$ & $J_{o,sub}$ & $O_{o,sub}$\\ \hline
EUHFORIA vs. MULTI--VP   & 0.95 & 0.98 & 0.93 & 0.97 & 0.81 & 0.93 \\
EUHFORIA vs. PSI--PFSS   & 0.98 & 0.99 & 0.98 & 0.99 & 0.94 & 0.99 \\
EUHFORIA vs. WSA         & 0.97 & 0.98 & 0.95 & 0.98 & 0.86 & 1.00 \\
MULTI--VP vs. PSI--PFSS  & 0.95 & 0.97 & 0.93 & 0.96 & 0.79 & 0.94 \\
MULTI--VP vs. WSA        & 0.96 & 0.98 & 0.95 & 0.98 & 0.87 & 0.96 \\
PSI--PFSS vs. WSA        & 0.96 & 0.98 & 0.94 & 0.97 & 0.82 & 0.99 \\ \hline

\hline
\multicolumn{7}{c}{2010--09--20} \\ \hline
 & $J_g$ & $O_g$ & $J_{g,sub}$ & $O_{g,sub}$ & $J_{o,sub}$ & $O_{o,sub}$\\ \hline
EUHFORIA vs. MULTI--VP   & 0.96 & 0.98 & 0.94 & 0.97 & 0.82 & 0.94 \\
EUHFORIA vs. PSI--PFSS   & 0.99 & 1.00 & 0.98 & 0.99 & 0.94 & 0.98 \\
EUHFORIA vs. WSA         & 0.97 & 0.99 & 0.96 & 0.98 & 0.87 & 0.99 \\
MULTI--VP vs. PSI--PFSS  & 0.96 & 0.98 & 0.94 & 0.97 & 0.81 & 0.95 \\
MULTI--VP vs. WSA        & 0.97 & 0.99 & 0.96 & 0.98 & 0.87 & 0.94 \\
PSI--PFSS vs. WSA        & 0.97 & 0.98 & 0.95 & 0.97 & 0.84 & 0.98 \\ \hline

\hline
\end{tabular}}
\end{table}

\begin{table}
\centering
\caption{Similarity metrics from mode-to-model comparisons of the open and closed field topologies simulated using the HMI ADAPT maps without ARs added retrospectively.}
\label{tab:open_closed_topology_comparisons_without}
\begin{tabular}{ l  c  c  c  c  c  c }
\hline

\hline
\multicolumn{7}{c}{2010--09--18} \\ \hline
 & $J_g$ & $O_g$ & $J_{g,sub}$ & $O_{g,sub}$ & $J_{o,sub}$ & $O_{o,sub}$\\ \hline
EUHFORIA vs. MULTI--VP   & 0.96 & 0.98 & 0.94 & 0.97 & 0.82 & 0.94 \\
EUHFORIA vs. PSI--PFSS   & 0.99 & 1.00 & 0.98 & 0.99 & 0.95 & 0.99 \\
EUHFORIA vs. WSA         & 0.97 & 0.99 & 0.95 & 0.98 & 0.87 & 0.99 \\
MULTI--VP vs. PSI--PFSS  & 0.95 & 0.98 & 0.93 & 0.96 & 0.80 & 0.94 \\
MULTI--VP vs. WSA        & 0.97 & 0.98 & 0.96 & 0.98 & 0.89 & 0.96 \\
PSI--PFSS vs. WSA        & 0.97 & 0.98 & 0.95 & 0.97 & 0.84 & 0.99 \\  \hline

\hline
\multicolumn{7}{c}{2010--09--19} \\ \hline
 & $J_g$ & $O_g$ & $J_{g,sub}$ & $O_{g,sub}$ & $J_{o,sub}$ & $O_{o,sub}$\\ \hline
EUHFORIA vs. MULTI--VP   & 0.96 & 0.98 & 0.93 & 0.96 & 0.79 & 0.93  \\ 
EUHFORIA vs. PSI--PFSS   & 0.99 & 0.99 & 0.99 & 0.99 & 0.95 & 0.99 \\ 
EUHFORIA vs. WSA         & 0.97 & 0.99 & 0.95 & 0.98 & 0.85 & 0.99 \\ 
MULTI--VP vs. PSI--PFSS  & 0.96 & 0.98 & 0.93 & 0.96 & 0.78 & 0.94  \\ 
MULTI--VP vs. WSA        & 0.97 & 0.98 & 0.95 & 0.98 & 0.85 & 0.94 \\ 
PSI--PFSS vs. WSA        & 0.97 & 0.98 & 0.95 & 0.97 & 0.84 & 1.00 \\  \hline

\hline
\multicolumn{7}{c}{2010--09--20} \\ \hline
 & $J_g$ & $O_g$ & $J_{g,sub}$ & $O_{g,sub}$ & $J_{o,sub}$ & $O_{o,sub}$\\ \hline
EUHFORIA vs. MULTI--VP   & 0.95 & 0.98 & 0.93 & 0.97 & 0.77 & 0.94  \\ 
EUHFORIA vs. PSI--PFSS   & 0.99 & 1.00 & 0.99 & 0.99 & 0.95 & 0.99 \\ 
EUHFORIA vs. WSA         & 0.97 & 0.99 & 0.95 & 0.98 & 0.83 & 0.98 \\ 
MULTI--VP vs. PSI--PFSS  & 0.95 & 0.98 & 0.93 & 0.96 & 0.76 & 0.95 \\ 
MULTI--VP vs. WSA        & 0.96 & 0.98 & 0.96 & 0.98 & 0.88 & 0.94 \\ 
PSI--PFSS vs. WSA        & 0.97 & 0.98 & 0.95 & 0.97 & 0.81 & 0.98 \\  \hline

\hline
\end{tabular}
\end{table}

\subsubsection{PFSS vs MHD modeled topology inter--comparison}
\label{PFSS_MHD_intercomp}

\begin{figure}[t!]
\centering
    \begin{minipage}{0.95\textwidth}
    \centering
        \includegraphics[width = 0.55\textwidth]{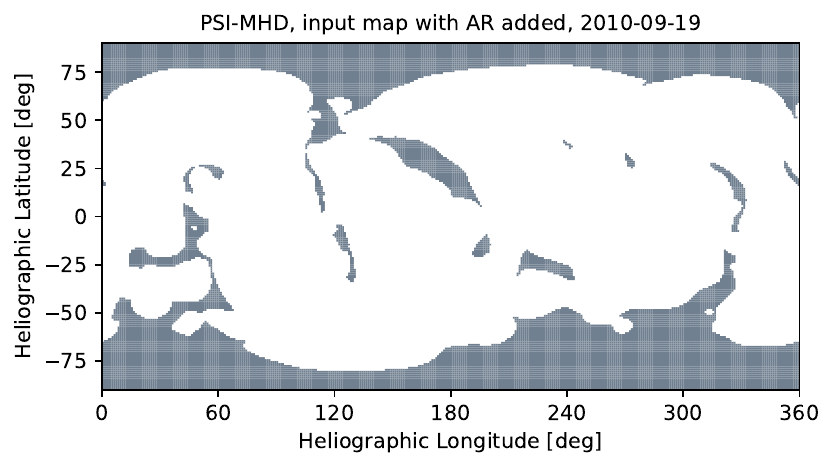}
    \end{minipage}
    \caption{Modeled  open (gray) and closed (white) magnetic field topology for the 2010--09--19 using the PSI MHD model described in Appendix~\ref{ap1:models} initiated with smoothed HMI ADAPT maps with ARs added retrospectively.}
    \label{fig:PSI_MHD}
\end{figure}

Considering the good agreement obtained between the open-closed field maps generated by the PFSS-based models, it is interesting to compare them to the one produced by the MHD model developed at PSI and described in Appendix~\ref{ap1:psi_models}. This is done for the date 2010--09--19. The modeled open and closed field topology produced by the MHD model was initiated with a smoothed HMI ADAPT global magnetic map for 2010--09--19 as an inner boundary condition, the same one also used for the PSI-PFSS runs analyzed in Section~\ref{smoothing_comparison}. The resulting open and closed field topology generated by the PSI-MHD model is shown in Figure~\ref{fig:PSI_MHD}. As discussed in Section~\ref{smoothing_comparison}, smoothing of the used global magnetic map leads to the loss of fine structures inside and around CHs. It can also lead to larger open field areas and the appearance of open field in more regions compared to a map generated by a simulation run initiated with a non-smoothed global magnetic map. It is thus, understandable that discrepancies between the MHD output maps and the ones by the PFSS models are expected. 

\begin{figure*}[t!]
\centering
\centering
    \begin{minipage}{0.95\textwidth}
        \includegraphics[width = 0.95\textwidth]{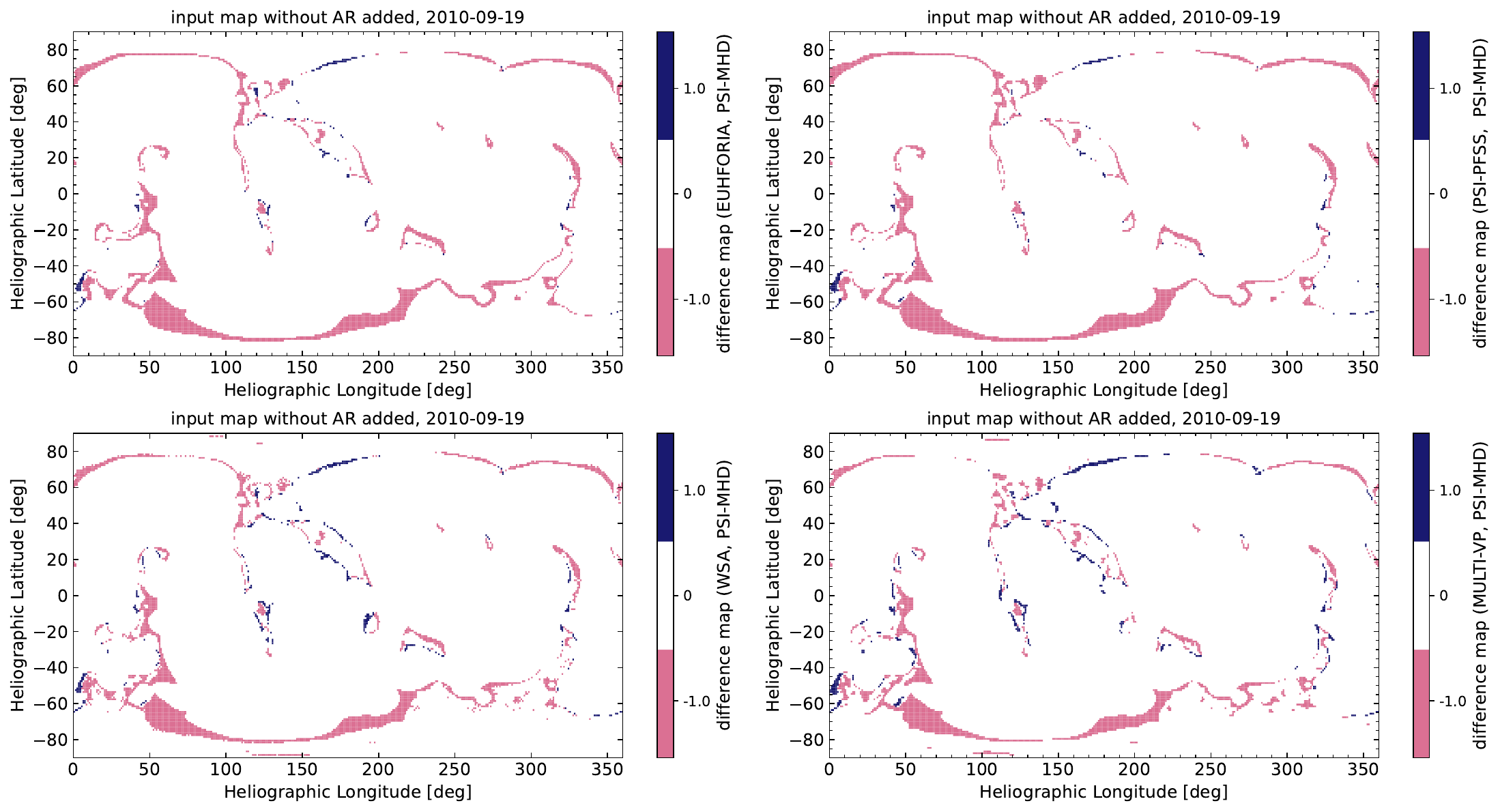}
    \end{minipage}
    \caption{Difference maps between the simulated open and closed field topologies by the PSI-MHD and each of the four PFSS models considered in this study in the case of being initiated with the HMI ADAPT maps with ARs added retrospectively for the 2010--09--19.}
    \label{fig:20100919_with_difference_maps_with_MHD}
\end{figure*}

Difference images between maps produced by each of the four PFSS models considered and the PSI-MHD one are given in Figure~\ref{fig:20100919_with_difference_maps_with_MHD}. All PFSS model outputs show differences to the PSI-MHD output throughout the entire Carrington map, and in particular at the south polar open field region. More precisely, in those regions, the PSI--MHD produces larger open field areas. This could not be fully attributed to the use of a smoothed global magnetic map for the PSI--MHD versus a non-smoothed for the others, because, as we showed in Section~\ref{smoothing_comparison}, smoothing does not significantly affect the modeled areas. The open field is determined by the intrinsic thermodynamics of the MHD model.

The calculated metrics for these comparisons are given in Table~\ref{tab:PFSS_to_MHD_comparisons}. The obtained values suggest a relatively fair agreement between the PFSS and MHD model outputs. For the case of $J_g$ and $O_g$ metrics this is due to the good agreement over closed field areas which as explained in earlier sections bias the metrics. When comparing the open field areas of the sub-maps the calculated $J_{o,sub}$ scores low, between 0.7--0.8, which implies that the open field area associated with the CH shows significant differences between a given PFSS and the PSI--MHD generated maps. By assessing the difference maps in Figure \ref{fig:20100919_with_difference_maps_with_MHD} one can see that the differences at the open field area of interest are concentrated on fine structures within the modeled CH area and at its boundary, similar to those shown in the bottom-left panel of Figure~\ref{fig:PSI_PFSS_smoothed}. We are therefore inclined to assume that the global magnetic map smoothing could be the source of some but not all of the discrepancies between MHD and PFSS model outputs.

\begin{table}
\centering
\caption{Calculated metrics for the similarity between the PFSS and the PSI--MHD simulation outputs generated with ADAPT maps with ARs added retrospectively.}\label{tab:PFSS_to_MHD_comparisons}
{\centering
\begin{tabular}{ l  c  c  c  c  c  c }
\hline

\hline
 & \hspace*{6mm} $J_g$ \hspace*{6mm} & $O_g$ & $J_{g,sub}$ & $O_{g,sub}$ & $J_{o,sub}$ & $O_{o,sub}$\\ \hline
EUHFORIA vs. PSI--MHD & 0.89 & 0.94 & 0.92 & 0.96 & 0.79 & 0.95 \\
MULTI--VP vs. PSI--MHD & 0.90 & 0.95 & 0.90 & 0.95 & 0.73 & 0.87 \\
PSI--PFSS vs. PSI--MHD & 0.89 & 0.94 & 0.93 & 0.96 & 0.79 & 0.97 \\
WSA vs. PSI--MHD & 0.90 & 0.95 & 0.92 & 0.96 & 0.80 & 0.89 \\ \hline

\hline
\end{tabular}}
\end{table}

\subsection{Comparison between modeled and observed CH area}
\label{Model_vs_observed_CH_area}

\begin{figure}
    \centering
     \includegraphics[width=0.85\textwidth]{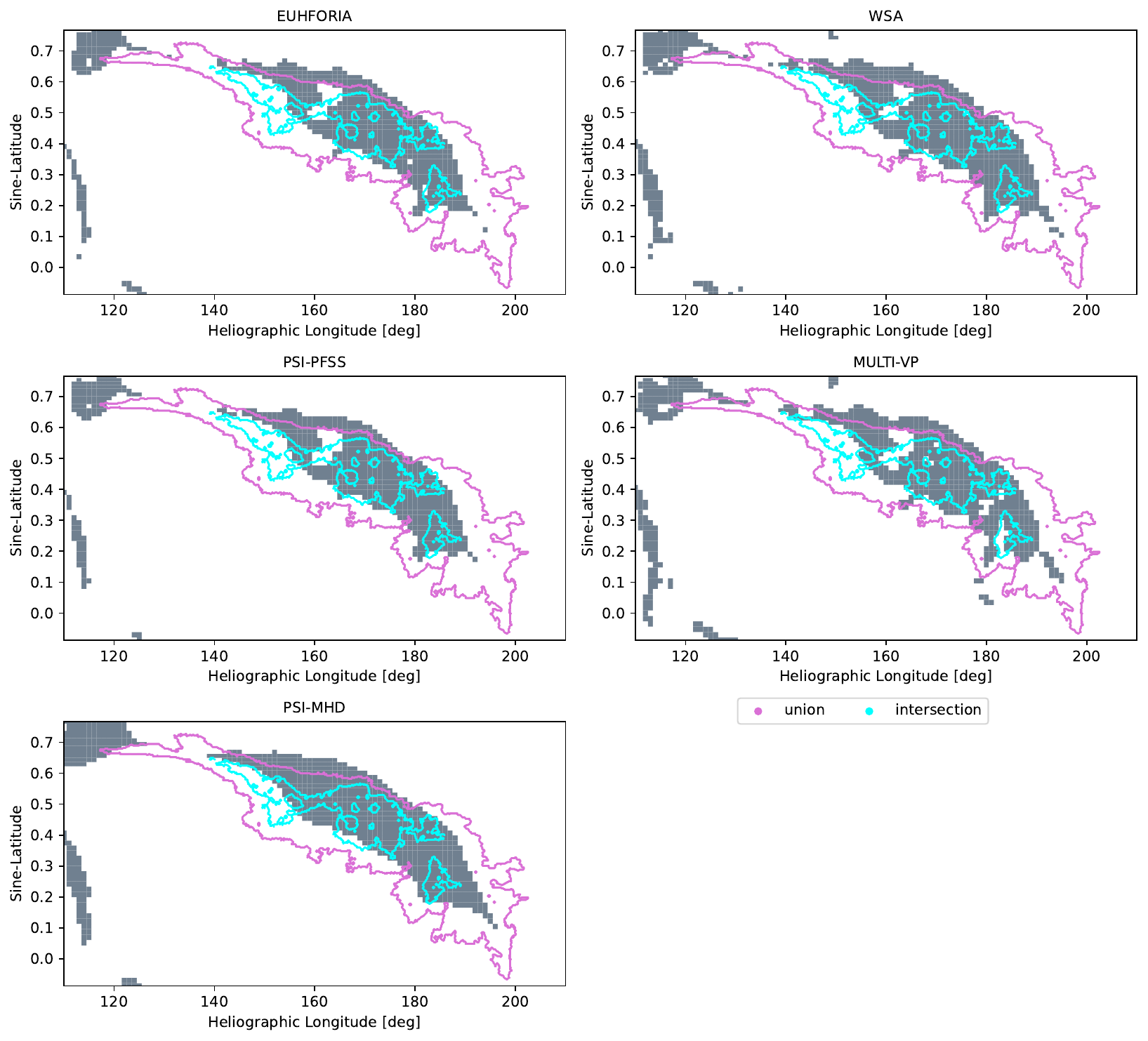}
    \caption{Modeled open and closed field topology at and near the under study CH using different models: EUHFORIA (top left), WSA (top right), PSI-PFSS (middle left), MULTI-VP (middle right), and PSI-MHD (bottom left). All models were initiated with the HMI ADAPT maps with ARs added retrospectively. CH boundaries extracted using CATCH on different input EUV filtergrams are over-plotted on each topology map, with magenta color indicating the union boundaries and cyan color the intersection boundaries (see Paper I for details on the union and intersection boundaries extraction).}
    \label{fig:CATCH_boundaries_overplotted}
\end{figure}

To validate the model outputs, we compare the simulated open field area that corresponds to the CH under study with the observed CH boundaries extracted in Paper I with CATCH. The models are here assessed by how well they capture the area associated with this CH using the metrics given in Equations~\ref{eq:jaccard_CH}, \ref{eq:overlap_CH}, and \ref{eq:coverage_CH}. Figure~\ref{fig:CATCH_boundaries_overplotted} shows the simulated open and closed field topology of the CH and its surrounding by the EUHFORIA (top-left), the WSA (top-right), the PSI--PFSS (middle-left), the MULTI--VP (middle-right), and PSI-MHD (bottom-left) models initiated by the HMI ADAPT global magnetic maps  with ARs added retrospectively for 2010--09--19. The same figure for the case of topologies initiated for the same date with the HMI ADAPT maps without ARs added retrospectively but without the PSI-MHD model output (which was not initiated) is given as supplementary material. This is the date when the CH was approximately centered at the central meridional zone seen from the Earth's field of view. Over-plotted in each panel are two CH boundaries from Paper I representing the union boundary (magenta outline) and the intersection boundary (cyan outline). We refer the reader to Paper I for more details on the boundaries. It is clear that for all models the simulated open field area does not cover the entire area enclosed by the magenta CH boundary. Namely, there is an abundance of closed field where observations indicate the field is likely to be open. Moreover, part of the modeled open field area extends beyond the CH boundary towards the northwest. When the cyan boundaries are considered, the modeled area greatly exceeds the CH boundary in the same direction. One can see that observed and modeled areas look quite different. With respect to observations this might be partly due to the global synchronic magnetic field dependence of CH structure, while the magnetic maps used in the model contain a lot of old and/or flux-evolved data. Another reason may be projection errors. For example when reprojecting a CH from one coordinate system to another the height of the CH in the respective line must be assumed. This may vary between 1\% -8\% above the solar surface. The difference between assumed and actual height of the CH location on the solar disk can have a significant effect due to the view angle.

To quantify this we calculate all the relevant metrics introduced in Section~\ref{sec:metrics}, $J_{CH}$, $O_{CH}$, and $cov$. These values are given in Table~\ref{tab:PFSS_to_CH}. The top section of the table is for the models initiated with the global magnetic map with ARs added retrospectively and the bottom section with the map without ARs. By definition, and more precisely due to their denominator, in some cases, the $O_{CH}$, and $cov$ are equal to each other. Overall when the modeled areas are compared to the union CATCH boundary the $J_{CH}$ obtains values below 0.5 with some cases scoring as low as 0.3 (see column 2 of the table). The main reason for the low score is due to the area marked by the union being larger than the modeled area. The opposite is true for the area marked with the intersection CATCH CH boundary, which is seemingly smaller than the modeled ones; however, this reversed area balance is still leading to low values of $J_{CH}$ which are listed in column 5. As the smallest area marks the denominator of $O_{CH}$ this leads to larger values for this metric as listed in columns 3 and 6 of the table. The coverage which indicates the portion of the area enclosed by the CATCH CH boundary and which was successfully modeled as open as expected scores moderately for comparisons to the largest CH boundary area marked by the CATCH union boundary ( see column 4--obtaining values between 0.4 and 0.6) versus comparisons to the smallest CH boundary area marked by the CATCH intersection boundary (see column 7). Simulation runs generated by the global magnetic map without ARs added retrospectively show similar patterns for the metrics given in the last four rows of the same table.

\begin{table}
    \centering
        \caption{Comparison of the observed CH on the 2010-09-19 with the PFSS modeled open field area associated with it using CATCH extracted boundaries from Paper I.}
        \label{tab:PFSS_to_CH}
        
        {\centering
        \begin{tabular}{ l  c  c  c || c  c  c }
        \hline
        
        \hline
        \multicolumn{7}{c}{Model output based on ADAPT maps with ARs added retrospectively}\\ \hline
         & \multicolumn{3}{c||}{Comparison with union} & \multicolumn{3}{c}{Comparison with intersection} \\
        & \multicolumn{3}{c||}{CATCH boundary} & \multicolumn{3}{c}{CATCH boundary} \\ \cline{2-7}
        & $J_{CH}$ & $O_{CH}$ & $cov$ & $J_{CH}$ & $O_{CH}$ & $cov$ \\ \hline
        EUHFORIA & 0.27 & 0.60 & 0.33 & 0.28 & 0.73 & 0.73 \\
        MULTI--VP & 0.28 & 0.54 & 0.37 & 0.24 & 0.73 & 0.73 \\
        PSI--PFSS & 0.26 & 0.62 & 0.31 & 0.28 & 0.72 & 0.72 \\
        PSI--PFSS smoothed & 0.29 & 0.59 & 0.36 & 0.27 & 0.77 & 0.77\\
        PSI--MHD & 0.30 & 0.56 & 0.40 & 0.26 & 0.82 & 0.82\\
        WSA & 0.30 & 0.56 & 0.39 & 0.25 & 0.77 & 0.77 \\ \hline
        
        \hline
        \multicolumn{7}{c}{Model output based on ADAPT maps without ARs added retrospectively}\\
        \hline
         & \multicolumn{3}{c||}{Comparison with union} & \multicolumn{3}{c}{Comparison with intersection} \\
        & \multicolumn{3}{c||}{CATCH boundary} & \multicolumn{3}{c}{CATCH boundary} \\ \cline{2-7}
        & $J_{CH}$ & $O_{CH}$ & $cov$ & $J_{CH}$ & $O_{CH}$ & $cov$ \\ \hline
        EUHFORIA & 0.23  & 0.54 & 0.29 & 0.27 & 0.69 & 0.69 \\
        MULTI--VP & 0.27 & 0.54 & 0.35 & 0.24 & 0.71 & 0.71 \\
        PSI--PFSS & 0.23 & 0.58 & 0.28 & 0.28 & 0.68 & 0.68 \\
        WSA & 0.26 & 0.53 & 0.33 & 0.26 & 0.75 & 0.75 \\ \hline

        \hline
        \end{tabular} \par }

\end{table}

\section{Discussion and Conclusions}
\label{sec_discussion}
In this study, we assess whether the selection of initiation data, the simulation set-up, and the employment of different numerical schemes resulted in comparable areas of open and closed magnetic field generated by implementations based on the same MHD approximation, namely the PFSS model. These models are the EUHFORIA coronal model, the MULTI-VP, the PSI-PFSS, and the WSA models. We focused our modeling efforts on three consecutive dates centered around 2010--09--19, the date during which the CH we studied in Paper I was located in the central meridional zone of the Sun as viewed from Earth. Our date selection was motivated by the decision to use the specific CH, and more precisely, the CH boundaries determined in Paper I, in order to validate the simulated open and closed field topology maps at the solar surface generated by the PFSS models. Further validation of these models was done using the output of the state-of-the-art PSI--MHD model. Our analysis focused both on comparing the entire Carrington map but also the specific CH area studied in Paper I.

We first considered the input global magnetic map and how that may affect the output magnetic field topology maps. We used two types of HMI ADAPT global magnetic maps; one with and one without ARs added retrospectively. Although the team which develops the ADAPT maps provided the most optimal out of the 12 realizations they generated, we also investigated the extent to which the topology maps initiated by each of the realizations differ from each other. This was done using only the WSA model to reduce the computational costs and simplify the analysis. From model intra-comparisons we deduced that all realizations produce highly comparable open and closed field topologies. Therefore, continuing the rest of our analysis with one realization only was justified.

Our next step was to compare whether adding ARs to the HMI ADAPT global magnetic maps retrospectively affects the modeled topologies. This is expected, since ARs, and more precisely their flux content, can influence both open field areas in their vicinity, but can also contribute to the global magnetic field topology of the solar corona. Due to our current observational limitations, we only have accurate and timely observations of a small sector of the global solar surface. Thus, information over the entire disc might be based on past or future observations and will not reflect well the influence of ARs present away from the central meridian as seen from Earth. Despite the anticipated differences between simulation outputs initiated by the two types of global magnetic maps, this was not reflected in our results. On the contrary, both types of maps generated comparable open and closed field topologies for all the PFSS models considered.

Quite commonly, pre-processing of the input global magnetic maps is employed, with smoothing being a widely used method by many coronal models. Smoothing will result in loss of magnetic variability, especially in cases where a heavy smoothing factor is applied, which will affect the local and global magnetic field topology output. Despite visible differences in our simulation outputs presented in Section~\ref{smoothing_comparison}, namely loss of fine structures within or around areas of open field, the maps compare relatively well to each other, with all the metrics considered scoring relatively high.

Our investigations surrounding the input global magnetic maps did not follow the path of previous studies focusing on assessing the effect on the simulation due to different magnetic field observations and techniques \citep{riley2014, lietal2021}. Our analysis considered predominantly our observational limitation and the  effect of outdated observations on ARs and how these translate to the simulation output. One aspect we did not address, but which is however interesting for future investigations, is the magnetograph saturation discussed in \citet[][]{wang22} for the MWO and the WSO, and how it affects modeling open field areas.

When it comes to simulation initiation, the source surface height, marking the outer boundary of the PFSS model, is another debatable parameter \citep[][]{lee11, ardenetal14, asvestari19, asvestari20, wagner22}. Coupled with the inner boundary of the SCS model, variations in their values have an immediate effect on the open field areas modeled. By using EUHFORIA and 5 different pairs of these two boundary heights we obtained the strongest differences between two modeled maps, which is in accordance with findings from previous works \citep[see for example][]{lee11, ardenetal14, asvestari19, asvestari20, wagner22}.

All the results listed so far in the conclusions section are based on comparisons of the global topology over the entire Carrington map (comparisons both for open and closed fields), but they are also supported by comparisons of sub-maps focused on the area of the CH of interest and its vicinity. The high values scored by the map-to-map comparisons are due to the agreement between two maps over the closed field areas. More precisely, closed field areas cover the majority of the maps and thus dominate the results. However, when applying the same analysis but only to the open field area associated with the specific CH, then all metrics -- although they still score adequately high -- give values below those for the global topology comparisons. As open coronal fields form the foundations for the interplanetary magnetic field and fast solar wind, it is important to specifically assess the models in these areas, in addition to global topology (open and closed field).

The model-to-model comparisons do not show much difference between all the PFSS models. However, discrepancies are found primarily at the boundary of the open field areas. The EUHFORIA coronal model and PSI-PFSS show the best agreement with each other compared to the other two models. The MULTI--VP simulates larger areas of open field in contrast to the other three models. Despite the dissimilarities between models visible in difference maps presented, the agreement between the PFSS models is of great interest. It suggests that different implementation methods generate comparable topologies. Validation of the PFSS models using the PSI-MHD simulation output shows that the latter generates significantly more open flux, in particular in the south polar region. These differences would most probably not be enough to account for the missing open flux problem. The fact that all models agree does not mean they are correct, but it does mean that at least we can have confidence in that the implementation issues can be ruled out. Validation of all the models using observed CH boundaries extracted with CATCH in Paper I showed the strong discrepancy between modeled and observed areas of open field. This is in accordance to earlier studies \citep[][]{asvestari19, asvestari20, wagner22}. The source of discrepancies can lie both with the model and the extraction methods from observations. To eliminate uncertainties introduced by the latter, we opted in using the previously studied CH boundaries from Paper I. A visible shift in the modeled open field area with respect to the CH area determined by the CATCH boundary could be of interest; however, earlier research has shown that shifts are not systematic \citep{asvestari20}. A larger sample could help further investigate the subject. It is noteworthy that while individual comparisons between observed coronal holes and modeled open field topologies (like the one in this paper) show big discrepancies, \citet{wallace19} showed that when the comparisons were made over 3 Carrington rotation time scales, the PFSS models and the observed coronal holes agreed well on average.

When comparing the global topology (both open and closed field pixels agreement) for two Carrington maps or two sub-maps the result is influenced by the closed field areas which occupy the largest portion of the map. Therefore, changes in the open field area match/miss-match will not have a strong effect on the overall map agreement and subsequently the metrics. This explains the high values the metrics obtain and highlights why it is important when focusing on a specific open field area to look at the agreement in open and closed field topologies and to also focus on how the maps agree in open field areas alone.

\section{Summary}
Our hypothesis in this study was that intra- and inter-comparisons will indicate differences among the modeled topologies that may be of significance, and need to be quantified. We anticipated that this will shed some light to the open flux problem and most certainly on the status of our current computationally inexpensive modeling. From our analysis it can be concluded that, despite their differences, models agree relatively well with each other in what they reconstruct as open and closed field. When comparing modeled open field associated with the CH under study, it could be shown that none of the models reproduced with reasonable accuracy the observed CH boundaries (see also Paper I). These add insights about missing open flux. Namely, the numerical methods and map processing is not the predominant source of the missing open flux, and thus new avenues are necessary to be investigated to address this open question. The discrepancies observed throughout the study, whether small or significant, are likely to be passed on to heliospheric models, increasing the variability/uncertainty of background solar wind and interplanetary magnetic field simulations, and subsequently, CME propagation modeling. This may increase the challenge for more accurate Space Weather forecasts.

We thank the International Space Science Institute (ISSI, Bern) for the generous support of the ISSI team “Magnetic open flux and solar wind structuring in interplanetary space” (2019-2022). 
The ADAPT model development is supported by Air Force Research Laboratory (AFRL), along with AFOSR (Air Force Office of Scientific Research) tasks 18RVCOR126 and 22RVCOR012. The views expressed are those of the authors and do not reflect the official guidance or position of the United States Government, the Department of Defense (DoD) or of the United States Air Force.
EA acknowledges support from the Academy of Finland/Research Council of Finland (Postdoctoral Researcher grant number 322455 and Academy Research Fellow grant number 355659). 
J.A.L. and R.M.C. were supported by the NASA Heliophysics Guest Investigator program (grants 
NNX17AB78G and 80NSSC19K0273), the NASA HSR program (grant 80NSSC18K0101 and 80NSSC18K1129), 
NASA LWS Strategic Capabilities  (grant 80NSSC22K0893), the NSF PREEVENTS program (grant ICER1854790)), and the STEREO SECCHI contract to NRL (under subcontract N00173-19-C-2003 to PSI).  Computational resources were provided by NASA's NAS (Pleiades) and NSF's XSEDE (TACC \& SDSC). 
M.M. acknowledges DFG-grant WI 3211/8-1 and is partly supported by the Bulgarian National Science Fund, grant No. KP-06-N44/2. 
SJH was supported, in part, by the NASA Heliophysics Living With a Star Science Program under Grant No. 80NSSC20K0183 and by the German Science Fund under Grant No. 448336908. 
SGH acknowledges funding by the Austrian Science Fund (FWF): Erwin-Schrödinger fellowship J-4560. 
C.N.A. is supported by the NASA competed Heliophysics Internal Scientist Funding Model (ISFM). 
MO is funded by part-funded by Science and Technology Facilities Council (STFC) grant numbers ST/R000921/1 and ST/V000497/1 and NERC grant number NE/S010033/1. 
C.S. acknowledges support from the NASA Living With a Star Jack Eddy Postdoctoral Fellowship Program, administered by UCAR's Cooperative Programs for the Advancement of Earth System Science (CPAESS) under award no. NNX16AK22G, and from NASA grants 80NSSC19K0914, 80NSSC20K0197, and 80NSSC20K0700. 
E.S. acknowledges support from a PhD grant awarded by the Royal Observatory of Belgium. E.S. research was supported by an appointment to the NASA Postdoctoral Program at the NASA Goddard Space Flight Center, administered by Oak Ridge Associated Universities under contract with NASA. 
JP acknowledges support from the SolMAG project (ERC-COG 724391) funded by the European Research Council (ERC) in the framework of the Horizon 2020 Research and Innovation Programme, the Finnish Centre of Excellence in Research of Sustainable Space (Academy of Finland grant number 312390), and the SWATCH project (Academy of Finland grant number 343581).

\appendix

\section{Detailed description of the employed models}
\label{ap1:models}

\subsubsection{WSA model: PFSS + SCS}
\label{sec:wsa_model}

The WSA model \citep[][]{arge00,arge03,arge04,mcgregor08,wallace19} uses ground-based observations of the Sun’s surface magnetic field, in the form of global photospheric magnetic maps, as its input. These maps are then used in a magnetostatic potential PFSS model \citep[][]{schatten69,altschuler69,wang92}, which solves Laplace’s equation using a spherical harmonic expansion to determine the coronal field out to the source surface, at 2.51~$R_{\odot}$. The output of the PFSS model serves as input to the Schatten Current Sheet (SCS) model \citep[][]{schatten71}, which provides a more realistic magnetic field topology of the upper corona. Only the innermost portion (i.e., from 2.49$R_{\odot}$ to between 5 and 30~$R_{\odot}$) of the SCS solution, which actually extends out to infinity, is used. A small overlap region between the PFSS and SCS is used to interface the two models, where the radial magnetic field components of the PFSS field solution at 2.49~$R_{\odot}$ is used as input to the SCS model. An empirical velocity relationship \citep[e.g.][]{arge03,arge04} is then used to assign solar wind speed at this outer boundary. It is a function of two coronal parameters the flux tube expansion factor ($f_s$) and the minimum angular separation ($\theta_b$) at the photosphere between an open field footpoint and the nearest coronal hole boundary. The model provides the radial magnetic field and solar wind speed at the outer coronal boundary surface, which is then fed into a simple, quick running 1-D kinematic solar wind model \citep[][]{arge00,arge04} or advanced three-dimensional (3D) MHD model solar wind propagation models such as MS-FLUKSS \citep[][]{kim20,singh20,manoharan15}, LFM-helio \citep[][]{pahud12,merkin16}, Gamera \citep[][]{zhang19}, and Enlil \citep[][]{odstrcil05,lee13,lee15,mcgregor11}. Densities and temperatures, which are not provided by WSA and required by MHD models, are deduced by assuming, for example, mass flux conservation and pressure balance \citep[e.g.,][]{kim20,merkin16}.

\subsubsection{EUHFORIA model: PFSS + SCS}
\label{sec:euhforia_model}

Similar to WSA, EUHFORIA \citep{pomoell18} consists of two modeling domains, the low corona that starts from the solar surface and extends up to the source surface at distance $R_{ss}$, and the upper corona with the inner boundary slightly below the source surface, at a height $R_{i}$, and its outer boundary at 0.1 AU (21.5~$R_{\odot}$). In the low corona, the magnetic field is computed following the PFSS approximation \citep[][]{altschuler69}, while in the upper coronal domain the employed model is the SCS model \citep[][]{schatten69, schatten71}. The default $[R_{i}, R_{ss}]$ pair of heights is the one proposed by \citet{mcgregor08}, namely, [2.3$R_{\odot}$, 2.6$R_{\odot}$]. To solve the Laplace equation for the PFSS and SCS models the selected numerical scheme is that of an expansion in solid harmonics. For the purpose of this paper the expansion was computed up to degree $l = 140$. Usually, a Gaussian smoothing can also be applied to the input global magnetic map with a $\sigma = 0.8$. For more details on the model architecture, we refer the reader to \citet{pomoell18}. For EUHFORIA the field lines are traced both from the bottom up and from the top down and the results are combined to built one map at the height of the solar surface.

\subsubsection{MULTI--VP model}
\label{sec:multivp_model}

The MULTI--VP model provides the radial profiles and amplitudes of slow and fast solar wind flows in a fraction or in the whole spherical domain. The simulation domain starts from the solar surface and extends typically up to 30~$R_{\odot}$. Being modular by construction, it offers the opportunity to combine different input global magnetic maps, coronal field extrapolation techniques, and coronal heating scenarios. For this study, MULTI--VP used the same global magnetic maps as the other considered models (see Section \ref{sec:magnetograms} for the details on the input maps), and two different types of extrapolation strategies were tested. Firstly, MULTI--VP was setup to use its own implementation of the PFSS technique using a standard source-surface radius of 2.5~$R_{\odot}$, from which the magnetic field is extended radially outwards. An additional correction is introduced in order to make the magnetic field in the high corona (above the source-surface) smoothly acquire an uniform-per-sector amplitude, as expected in the interplanetary medium. This correction operates within the first 12~$R_{\odot}$ of the domain and consists of small adjustments to the magnetic field expansion ratio (that change the PFSS superradial expansion rates by a maximum factor of 2, in the most extreme cases). Total open flux is not affected by this transformation. Alternatively, MULTI--VP was also setup to use directly the PFSS+SCS  extrapolations provided by the WSA model (Section \ref{sec:wsa_model}), thus skipping its native extrapolation stage. This approach was followed for the open closed field maps presented in this paper. The PFSS+SCS field line tracings are, however, reprocessed in order to eliminate sharp boundaries at the transition between the PFSS and the SCS parts of the domain and at the lower boundary (if they ever occur). This is a requirement for the subsequent computation of the solar wind density, speed and temperature. The total open flux is not expected to be modified by this transformation.

\subsubsection{PSI MHD and PFSS models}
\label{ap1:psi_models}

Two different models developed at PSI are considered in this study, an MHD (PSI--MHD) and a PFSS (PSI--PFSS) model. For the PSI--MHD model, we employ a thermodynamic MHD simulation of the solar corona for the Carrington rotation (CR) 2101, performed with the Magnetohydrodynamic Algorithm outside a Sphere (MAS) code, The method of solution has been described previously \citep[e.g.,][]{mikiclinker94,linkeretal1999,lionelloetal1998,lionelloetal2009}.  Paper~1 employed a MAS simulation for this CR, with the boundary condition derived from an HMI LOS synoptic map for CR2101.  To more readily compare the MHD results in this paper with the other models, we performed a new simulation using an ADAPT-HMI realization (see section \ref{magnetograms} for the boundary condition.  This simulation was performed with one of the standard CORHEL coronal heating models (heating model 2) used for runs on the PSI website (e.g.  blue{\url{https://www.predsci.com/}}.  This heating model has a different functional form but similar properties to that described in \citet{lionelloetal2009}.  The simulation utilized a $255 \times 180 \times 360$ nonuniform $r,\theta,\phi$ mesh, with the smallest radial cells of  $\Delta r = 0.000243 R_S$ ($\approx 169$ km) near the solar surface and $\Delta r = 0.42R_S$ at the outer boundary of $21.5 R_S$.  For the co-latitude mesh, $\Delta \theta = 0.7^\circ$ near the equator and $1.7^\circ$ near the poles, and for the longitudinal mesh, $\Delta \phi =  1^\circ$ was uniform.  

For the PSI--PFSS, the PFSS solutions were performed using POT3D \citep{caplanetal2021}. The POT3D solves Laplace's equation using finite differences on a non-uniform logically-rectangular spherical grid. It is highly efficient and can be deployed on massively parallel and GPU-accelerated computers. It has recently been released as an open source project on GitHub~\footnote{\url{github.com/predsci/POT3D}}.

\section{Metrics definitions}
\label{ap2:metrics}

\subsubsection{Jaccard similarity coefficient}

The first metric we employ in this paper is the Jaccard similarity coefficient, also known as the Jaccard index, which quantifies the similarity and difference between two samples. It is the fraction of the intersection over the union of two samples A and B and it is given by:

\begin{equation}
(A,B) = \frac{\mathopen|A\cap B \mathclose|}{\mathopen|A \cup B \mathclose|},
\end{equation}

\noindent where $J(A,B)$ can take values between [0,1] with 1 occurring for perfect agreement between the two samples.
In quantifying the agreement over the whole modeled solar surface, A and B represent the two maps we compare to each other and the intersection is the total number of matching modeled open and closed field pixels, $N(A\equiv B)$, for example, areas where both maps are simultaneously open or simultaneously closed. We refer to this as the Jaccard global, $J_g$, and it is thus given by the formula:

\begin{equation} \label{eq:jaccard_global}
J_g = \frac{N(A\equiv B)}{N_{A} + N_{B} - N(A\equiv B)},
\end{equation}

\noindent where $N_{A}$ and $N_{B}$ are the total sizes of A and B (as in total number of pixels of the grid). Considering that the two maps are on an identical grid, $N_{A} = N_{B}$.

Besides comparing the whole solar surface maps we also focus on the specific area that encompasses the CH of interested and which we also studied in Paper~I. In this case we produce the same area sub-maps for each model output and we quantify the Jaccard global for each pair of sub-maps, $J_{g,sub}$, which is given by the formula:

\begin{equation} \label{jaccard_global_sub}
J_{g,sub} = \frac{N(sub(A)\equiv sub(B))}{N_{sub(A)} + N_{sub(B)} - N(sub(A)\equiv sub(B))},
\end{equation}

\noindent where $N(sub(A)\equiv sub(B))$ is the total number of pixels for which the two sub-maps agree (both for closed and open field), while $N_{sub(A)}$ and $N_{sub(A)}$ are the total sizes of the two sub-maps. For our analysis we always select the two sub-maps to be the identical grid, therefore, $N_{sub(A)} = N_{sub(A)}$.

Because the closed field areas compose the largest part of the solar surface map, they greatly influence the calculated $J_{g}$ and $J_{g,sub}$ metrics, with the latter being inclined to obtain values closer to 1. Since the differences between two maps are more apparent at the open field areas we calculate what we refer to as the sub-map Jaccard open, $J_{o,sub}$. This metric is estimated in addition to the $J_{g,sub}$. For $J_{o,sub}$ we did not to consider matching pixels of closed flux between the two sub-maps and since the open field area modeled in sub-map A can differ in size from that modeled in sub-map B the metric will be given by the formula:

\begin{equation} \label{jaccard_open_sub}
J_{o,sub} = \frac{N_{open}(sub(A)\equiv sub(B))}{N_{open,sub(A)}+ N_{open,sub(B)}-N_{open}(sub(A)\equiv sub(B))},
\end{equation}

\noindent where $N_{open}(sub(A)\equiv sub(B))$ is the number of open pixels that match on both sub-maps, and $N_{open,sub(A)}$ and $N_{open,sub(B)}$ are the total number of open flux pixels on each sub-map A and B respectively. Note that the open field areas of the two sub-maps may not be of identical size, therefore $N_{open,sub(A)}$ and $N_{open,sub(B)}$ may not be equal, as the equivalent quantities in Equations \ref{eq:jaccard_global} and \ref{jaccard_global_sub} were. This Jaccard open we define here is the $P_{Jac}$ also used in \citet{wagner22}.

Besides model-model output topology comparisons we also assessed the output of each model by comparing it to the observed EUV CH topology. In this case, we wanted to evaluate what percentage of the observed CH in EUV was successfully modeled as an open field by a given global magnetic map-model pair. The modeled open field area corresponding to the CH of interest, described in Section \ref{sec:obs_CH}, is referred to as modeled CH in the manuscript. In this case the Jaccard similarity coefficient becomes:

\begin{equation} \label{eq:jaccard_CH}
J_{CH} = \frac{TP}{TP + FP + FN},
\end{equation}

\noindent where TP is the number of open field pixels of the modeled CH area that lay within the CATCH boundary, FP is the number of open field pixels of the modeled CH that lay outside the CATCH boundary, and FN are the modeled closed field pixels that lay within the CATCH boundary.  

\subsubsection{Overlap coefficient}
The second metric we consider is the overlap coefficient, also known as Szymkiewicz–Simpson coefficient, which is defined as the intersection of two samples A and B divided by the smallest size of the two samples, as shown in the formula below 

\begin{equation}
O(A,B) = \frac{\mathopen|A\cap B \mathclose|}{min(\mathopen|A\mathclose|, \mathopen|B\mathclose|)}, 
\end{equation}

\noindent similar to $J(A,B)$, $O(A,B)$ also take values between [0,1] with 1 occurring for absolute agreement of samples A and B.

In the case of the two samples A and B being identical then $\mathopen|A\mathclose| = \mathopen|B\mathclose|$ and thus $min(\mathopen|A\mathclose|, \mathopen|B\mathclose|) = \mathopen|A\mathclose| = \mathopen|B\mathclose|$. Subsequently, the Overlap coefficient when comparing the full solar surface maps or the sub-maps both for the open and closed field areas similarity becomes:

\begin{equation} \label{overlap_global}
O_g = \frac{N(A\equiv B)}{\mathopen|A\mathclose|}, 
\end{equation}

\noindent and

\begin{equation} \label{overlap_global_sub}
O_{g,sub} = \frac{N(sub(A)\equiv sub(B))}{\mathopen|sub(A)\mathclose|}, 
\end{equation}

\noindent respectively. However, if we do not include the similarity in the closed field for the sub-maps and only take into consideration the open field area then the overlap coefficient is:

\begin{equation}
O_{o,sub} = \frac{N_{open}(sub(A)\equiv sub(B))}{min(\mathopen|N_{open,sub(A)}\mathclose|,\mathopen|N_{open,sub(B)}\mathclose|)}.
\end{equation}

According to its definition, the $O_{o,sub}$ metric is higher when the smallest of two open field areas compared is a subset of the larger one, and thus, it will score higher than the equivalent $J_{o,sub}$.

In comparing the modeled CH area to the observed EUV boundaries extracted with CATCH, the Overlap coefficient becomes:

\begin{equation} \label{eq:overlap_CH}
O_{CH} = \frac{TP}{min(\mathopen|N_{mCH}\mathclose|, \mathopen|N_{oCH}\mathclose|)}, 
\end{equation}

\noindent where $N_{mCH}$ is the total number of pixels of the modeled CH open field pixels and $N_{oCH}$ is the total number of modeled pixels enclosed by the CATCH boundary, which are modeled either as open (true positives) or closed (false negative) field pixels.

\subsubsection{Coverage}

In addition to $J_{CH}$ and $O_{CH}$, for determining how well the models reproduce the area associated with the CH, we used the Coverage parameter introduced in \citet{asvestari19} and which is given by the following formula:

\begin{equation} \label{eq:coverage_CH}
cov = \frac{N_{open}}{N_{oCH}},
\end{equation}

where $N_{open}$ is the number of pixels on a simulation output map that are modeled as open field pixels and which are contained within the observed EUV CH boundaries (successfully modeled pixels -- true positives). Since coverage is unique to the EUV CH boundary, it is only applied to modeled-EUV CH comparisons and not for model-model comparisons. It is important to note that this parameter does not consider whether the modeled area is shifted with respect to the EUV CH boundary or whether the modeled area is larger than the observed EUV CH boundary.


\bibliography{final_bib}{}

\begin{thebibliography}{}
\expandafter\ifx\csname natexlab\endcsname\relax\def\natexlab#1{#1}\fi
\providecommand{\url}[1]{\href{#1}{#1}}
\providecommand{\dodoi}[1]{doi:~\href{http://doi.org/#1}{\nolinkurl{#1}}}
\providecommand{\doeprint}[1]{\href{http://ascl.net/#1}{\nolinkurl{http://ascl.net/#1}}}
\providecommand{\doarXiv}[1]{\href{https://arxiv.org/abs/#1}{\nolinkurl{https://arxiv.org/abs/#1}}}

\bibitem[{{Altschuler} \& {Newkirk}(1969)}]{altschuler69}
{Altschuler}, M.~D., \& {Newkirk}, G. 1969, \solphys, 9, 131,
  \dodoi{10.1007/BF00145734}

\bibitem[{{Antiochos} {et~al.}(2007){Antiochos}, {DeVore}, {Karpen}, \&
  {Miki{\'c}}}]{antiochos07}
{Antiochos}, S.~K., {DeVore}, C.~R., {Karpen}, J.~T., \& {Miki{\'c}}, Z. 2007,
  \apj, 671, 936, \dodoi{10.1086/522489}

\bibitem[{{Antiochos} {et~al.}(2011){Antiochos}, {Miki{\'c}}, {Titov},
  {Lionello}, \& {Linker}}]{antiochos11}
{Antiochos}, S.~K., {Miki{\'c}}, Z., {Titov}, V.~S., {Lionello}, R., \&
  {Linker}, J.~A. 2011, \apj, 731, 112, \dodoi{10.1088/0004-637X/731/2/112}

\bibitem[{{Arden} {et~al.}(2014){Arden}, {Norton}, \& {Sun}}]{ardenetal14}
{Arden}, W.~M., {Norton}, A.~A., \& {Sun}, X. 2014, Journal of Geophysical
  Research (Space Physics), 119, 1476, \dodoi{10.1002/2013JA019464}

\bibitem[{{Arge} {et~al.}(2013){Arge}, {Henney}, {Hernandez}, {Toussaint},
  {Koller}, \& {Godinez}}]{arge13}
{Arge}, C.~N., {Henney}, C.~J., {Hernandez}, I.~G., {et~al.} 2013, in American
  Institute of Physics Conference Series, Vol. 1539, Solar Wind 13, ed. G.~P.
  {Zank}, J.~{Borovsky}, R.~{Bruno}, J.~{Cirtain}, S.~{Cranmer}, H.~{Elliott},
  J.~{Giacalone}, W.~{Gonzalez}, G.~{Li}, E.~{Marsch}, E.~{Moebius},
  N.~{Pogorelov}, J.~{Spann}, \& O.~{Verkhoglyadova}, 11--14,
  \dodoi{10.1063/1.4810977}

\bibitem[{{Arge} {et~al.}(2010){Arge}, {Henney}, {Koller}, {Compeau}, {Young},
  {MacKenzie}, {Fay}, \& {Harvey}}]{Arge10}
{Arge}, C.~N., {Henney}, C.~J., {Koller}, J., {et~al.} 2010, in American
  Institute of Physics Conference Series, Vol. 1216, Twelfth International
  Solar Wind Conference, ed. M.~{Maksimovic}, K.~{Issautier},
  N.~{Meyer-Vernet}, M.~{Moncuquet}, \& F.~{Pantellini}, 343--346,
  \dodoi{10.1063/1.3395870}

\bibitem[{{Arge} {et~al.}(2004){Arge}, {Luhmann}, {Odstrcil}, {Schrijver}, \&
  {Li}}]{arge04}
{Arge}, C.~N., {Luhmann}, J.~G., {Odstrcil}, D., {Schrijver}, C.~J., \& {Li},
  Y. 2004, Journal of Atmospheric and Solar-Terrestrial Physics, 66, 1295,
  \dodoi{10.1016/j.jastp.2004.03.018}

\bibitem[{Arge {et~al.}(2003)Arge, Odstrcil, Pizzo, \& Mayer}]{arge03}
Arge, C.~N., Odstrcil, D., Pizzo, V.~J., \& Mayer, L.~R. 2003, AIP Conference
  Proceedings, 679, 190, \dodoi{10.1063/1.1618574}

\bibitem[{{Arge} \& {Pizzo}(2000)}]{arge00}
{Arge}, C.~N., \& {Pizzo}, V.~J. 2000, \jgr, 105, 10465,
  \dodoi{10.1029/1999JA000262}

\bibitem[{{Asvestari} {et~al.}(2019){Asvestari}, {Heinemann}, {Temmer},
  {Pomoell}, {Kilpua}, {Magdalenic}, \& {Poedts}}]{asvestari19}
{Asvestari}, E., {Heinemann}, S.~G., {Temmer}, M., {et~al.} 2019, Journal of
  Geophysical Research (Space Physics), 124, 8280, \dodoi{10.1029/2019JA027173}

\bibitem[{{Asvestari} {et~al.}(2020){Asvestari}, {Heinemann}, {Temmer},
  {Pomoell}, {Kilpua}, {Magdalenic}, \& {Poedts}}]{asvestari20}
{Asvestari}, E., {Heinemann}, S.~G., {Temmer}, M., {et~al.} 2020, in Journal of
  Physics Conference Series, Vol. 1548, Journal of Physics Conference Series,
  012004, \dodoi{10.1088/1742-6596/1548/1/012004}

\bibitem[{{Badman} {et~al.}(2020){Badman}, {Bale}, {Mart{\'\i}nez Oliveros},
  {Panasenco}, {Velli}, {Stansby}, {Buitrago-Casas}, {R{\'e}ville}, {Bonnell},
  {Case}, {Dudok de Wit}, {Goetz}, {Harvey}, {Kasper}, {Korreck}, {Larson},
  {Livi}, {MacDowall}, {Malaspina}, {Pulupa}, {Stevens}, \&
  {Whittlesey}}]{badman20}
{Badman}, S.~T., {Bale}, S.~D., {Mart{\'\i}nez Oliveros}, J.~C., {et~al.} 2020,
  \apjs, 246, 23, \dodoi{10.3847/1538-4365/ab4da7}

\bibitem[{{Barnes} {et~al.}(2023){Barnes}, {DeRosa}, {Jones}, {Arge}, {Henney},
  \& {Cheung}}]{barnes2023}
{Barnes}, G., {DeRosa}, M.~L., {Jones}, S.~I., {et~al.} 2023, \apj, 946, 105,
  \dodoi{10.3847/1538-4357/acba8e}

\bibitem[{{Caplan} {et~al.}(2021){Caplan}, {Downs}, {Linker}, \&
  {Mikic}}]{caplanetal2021}
{Caplan}, R.~M., {Downs}, C., {Linker}, J.~A., \& {Mikic}, Z. 2021, \apj, 915,
  44, \dodoi{10.3847/1538-4357/abfd2f}

\bibitem[{{Cohen} {et~al.}(2007){Cohen}, {Sokolov}, {Roussev}, {Arge},
  {Manchester}, {Gombosi}, {Frazin}, {Park}, {Butala}, {Kamalabadi}, \&
  {Velli}}]{cohenetal07}
{Cohen}, O., {Sokolov}, I.~V., {Roussev}, I.~I., {et~al.} 2007, \apjl, 654,
  L163, \dodoi{10.1086/511154}

\bibitem[{{Cranmer} {et~al.}(2017){Cranmer}, {Gibson}, \&
  {Riley}}]{cranmer2017}
{Cranmer}, S.~R., {Gibson}, S.~E., \& {Riley}, P. 2017, \ssr, 212, 1345,
  \dodoi{10.1007/s11214-017-0416-y}

\bibitem[{{Evans} {et~al.}(2012){Evans}, {Opher}, {Oran}, {van der Holst},
  {Sokolov}, {Frazin}, {Gombosi}, \& {V{\'a}squez}}]{evansetal12}
{Evans}, R.~M., {Opher}, M., {Oran}, R., {et~al.} 2012, \apj, 756, 155,
  \dodoi{10.1088/0004-637X/756/2/155}

\bibitem[{{Feng}(2020)}]{feng2020}
{Feng}, X. 2020, {Magnetohydrodynamic Modeling of the Solar Corona and
  Heliosphere}, \dodoi{10.1007/978-981-13-9081-4}

\bibitem[{{Fisk} \& {Zurbuchen}(2006)}]{fisk2006}
{Fisk}, L.~A., \& {Zurbuchen}, T.~H. 2006, Journal of Geophysical Research
  (Space Physics), 111, A09115, \dodoi{10.1029/2005JA011575}

\bibitem[{{Frost} {et~al.}(2022){Frost}, {Owens}, {Macneil}, \&
  {Lockwood}}]{Frostetal22}
{Frost}, A.~M., {Owens}, M., {Macneil}, A., \& {Lockwood}, M. 2022, \solphys,
  297, 82, \dodoi{10.1007/s11207-022-02004-6}

\bibitem[{{Groth} {et~al.}(2000){Groth}, {De Zeeuw}, {Gombosi}, \&
  {Powell}}]{grothetal00}
{Groth}, C. P.~T., {De Zeeuw}, D.~L., {Gombosi}, T.~I., \& {Powell}, K.~G.
  2000, \jgr, 105, 25053, \dodoi{10.1029/2000JA900093}

\bibitem[{{Guo} {et~al.}(2016){Guo}, {Xia}, {Keppens}, \& {Valori}}]{guoetal16}
{Guo}, Y., {Xia}, C., {Keppens}, R., \& {Valori}, G. 2016, \apj, 828, 82,
  \dodoi{10.3847/0004-637X/828/2/82}

\bibitem[{{Harra} {et~al.}(2021){Harra}, {Andretta}, {Appourchaux}, {Baudin},
  {Bellot-Rubio}, {Birch}, {Boumier}, {Cameron}, {Carlsson}, {Corbard},
  {Davies}, {Fazakerley}, {Fineschi}, {Finsterle}, {Gizon}, {Harrison},
  {Hassler}, {Leibacher}, {Liewer}, {Macdonald}, {Maksimovic}, {Murphy},
  {Naletto}, {Nigro}, {Owen}, {Mart{\'\i}nez-Pillet}, {Rochus}, {Romoli},
  {Sekii}, {Spadaro}, {Veronig}, \& {Schmutz}}]{Harraetal21}
{Harra}, L., {Andretta}, V., {Appourchaux}, T., {et~al.} 2021, Experimental
  Astronomy, \dodoi{10.1007/s10686-021-09769-x}

\bibitem[{{Hayashi} {et~al.}(2016){Hayashi}, {Yang}, \& {Deng}}]{Hayashi2016}
{Hayashi}, K., {Yang}, S., \& {Deng}, Y. 2016, Journal of Geophysical Research
  (Space Physics), 121, 1046, \dodoi{10.1002/2015JA021757}

\bibitem[{{Heinemann} {et~al.}(2018){Heinemann}, {Hofmeister}, {Veronig}, \&
  {Temmer}}]{heinemann18}
{Heinemann}, S.~G., {Hofmeister}, S.~J., {Veronig}, A.~M., \& {Temmer}, M.
  2018, \apj, 863, 29, \dodoi{10.3847/1538-4357/aad095}

\bibitem[{{Heinemann} {et~al.}(2019){Heinemann}, {Temmer}, {Heinemann},
  {Dissauer}, {Samara}, {Jer{\v{c}}i{\'c}}, {Hofmeister}, \&
  {Veronig}}]{heinemann19}
{Heinemann}, S.~G., {Temmer}, M., {Heinemann}, N., {et~al.} 2019, \solphys,
  294, 144, \dodoi{10.1007/s11207-019-1539-y}

\bibitem[{{Hickmann} {et~al.}(2015){Hickmann}, {Godinez}, {Henney}, \&
  {Arge}}]{Hickmann2015}
{Hickmann}, K.~S., {Godinez}, H.~C., {Henney}, C.~J., \& {Arge}, C.~N. 2015,
  \solphys, 290, 1105, \dodoi{10.1007/s11207-015-0666-3}

\bibitem[{{Hofmeister} {et~al.}(2019){Hofmeister}, {Utz}, {Heinemann},
  {Veronig}, \& {Temmer}}]{Hofmeister_etal_2019}
{Hofmeister}, S.~J., {Utz}, D., {Heinemann}, S.~G., {Veronig}, A., \& {Temmer},
  M. 2019, \aap, 629, A22, \dodoi{10.1051/0004-6361/201935918}

\bibitem[{{Hofmeister} {et~al.}(2017){Hofmeister}, {Veronig}, {Reiss},
  {Temmer}, {Vennerstrom}, {Vr{\v{s}}nak}, \& {Heber}}]{Hofmeister_etal_2017}
{Hofmeister}, S.~J., {Veronig}, A., {Reiss}, M.~A., {et~al.} 2017, \apj, 835,
  268, \dodoi{10.3847/1538-4357/835/2/268}

\bibitem[{{Jiang} {et~al.}(2012){Jiang}, {Feng}, \& {Xiang}}]{jiangetal12}
{Jiang}, C., {Feng}, X., \& {Xiang}, C. 2012, \apj, 755, 62,
  \dodoi{10.1088/0004-637X/755/1/62}

\bibitem[{{Karna} {et~al.}(2014){Karna}, {Hess Webber}, \& {Pesnell}}]{karna14}
{Karna}, N., {Hess Webber}, S.~A., \& {Pesnell}, W.~D. 2014, \solphys, 289,
  3381, \dodoi{10.1007/s11207-014-0541-7}

\bibitem[{{Kim} {et~al.}(2020){Kim}, {Pogorelov}, {Arge}, {Henney},
  {Jones-Mecholsky}, {Smith}, {Bale}, {Bonnell}, {Dudok de Wit}, {Goetz},
  {Harvey}, {MacDowall}, {Malaspina}, {Pulupa}, {Kasper}, {Korreck}, {Stevens},
  {Case}, {Whittlesey}, {Livi}, {Larson}, {Klein}, \& {Zank}}]{kim20}
{Kim}, T.~K., {Pogorelov}, N.~V., {Arge}, C.~N., {et~al.} 2020, \apjs, 246, 40,
  \dodoi{10.3847/1538-4365/ab58c9}

\bibitem[{{Kruse} {et~al.}(2021){Kruse}, {Heidrich-Meisner}, \&
  {Wimmer-Schweingruber}}]{kruseetal21}
{Kruse}, M., {Heidrich-Meisner}, V., \& {Wimmer-Schweingruber}, R.~F. 2021,
  \aap, 645, A83, \dodoi{10.1051/0004-6361/202039120}

\bibitem[{{Kruse} {et~al.}(2020){Kruse}, {Heidrich-Meisner},
  {Wimmer-Schweingruber}, \& {Hauptmann}}]{kruseetal20}
{Kruse}, M., {Heidrich-Meisner}, V., {Wimmer-Schweingruber}, R.~F., \&
  {Hauptmann}, M. 2020, \aap, 638, A109, \dodoi{10.1051/0004-6361/202037734}

\bibitem[{{Lee} {et~al.}(2015){Lee}, {Arge}, {Odstrcil}, {Millward}, {Pizzo},
  \& {Lugaz}}]{lee15}
{Lee}, C.~O., {Arge}, C.~N., {Odstrcil}, D., {et~al.} 2015, \solphys, 290,
  1207, \dodoi{10.1007/s11207-015-0667-2}

\bibitem[{{Lee} {et~al.}(2013){Lee}, {Arge}, {Odstr{\v{c}}il}, {Millward},
  {Pizzo}, {Quinn}, \& {Henney}}]{lee13}
{Lee}, C.~O., {Arge}, C.~N., {Odstr{\v{c}}il}, D., {et~al.} 2013, \solphys,
  285, 349, \dodoi{10.1007/s11207-012-9980-1}

\bibitem[{{Lee} {et~al.}(2011){Lee}, {Luhmann}, {Hoeksema}, {Sun}, {Arge}, \&
  {de Pater}}]{lee11}
{Lee}, C.~O., {Luhmann}, J.~G., {Hoeksema}, J.~T., {et~al.} 2011, \solphys,
  269, 367, \dodoi{10.1007/s11207-010-9699-9}

\bibitem[{{Levine} {et~al.}(1977){Levine}, {Altschuler}, \&
  {Harvey}}]{levine77}
{Levine}, R.~H., {Altschuler}, M.~D., \& {Harvey}, J.~W. 1977, \jgr, 82, 1061,
  \dodoi{10.1029/JA082i007p01061}

\bibitem[{{Levine} {et~al.}(1982){Levine}, {Schulz}, \&
  {Frazier}}]{levineetal82}
{Levine}, R.~H., {Schulz}, M., \& {Frazier}, E.~N. 1982, \solphys, 77, 363,
  \dodoi{10.1007/BF00156118}

\bibitem[{{Li} {et~al.}(2021){Li}, {Feng}, \& {Wei}}]{lietal2021}
{Li}, H., {Feng}, X., \& {Wei}, F. 2021, Journal of Geophysical Research (Space
  Physics), 126, e28870, \dodoi{10.1029/2020JA028870}

\bibitem[{{Linker} {et~al.}(1999){Linker}, {Miki{\'c}}, {Biesecker}, {Forsyth},
  {Gibson}, {Lazarus}, {Lecinski}, {Riley}, {Szabo}, \&
  {Thompson}}]{linkeretal1999}
{Linker}, J.~A., {Miki{\'c}}, Z., {Biesecker}, D.~A., {et~al.} 1999, \jgr, 104,
  9809, \dodoi{10.1029/1998JA900159}

\bibitem[{{Linker} {et~al.}(2017){Linker}, {Caplan}, {Downs}, {Riley}, {Mikic},
  {Lionello}, {Henney}, {Arge}, {Liu}, {Derosa}, {Yeates}, \&
  {Owens}}]{linker2017}
{Linker}, J.~A., {Caplan}, R.~M., {Downs}, C., {et~al.} 2017, \apj, 848, 70,
  \dodoi{10.3847/1538-4357/aa8a70}

\bibitem[{{Linker} {et~al.}(2021){Linker}, {Heinemann}, {Temmer}, {Owens},
  {Caplan}, {Arge}, {Asvestari}, {Delouille}, {Downs}, {Hofmeister}, {Jebaraj},
  {Madjarska}, {Pinto}, {Pomoell}, {Samara}, {Scolini}, \&
  {Vr{\v{s}}nak}}]{linker21}
{Linker}, J.~A., {Heinemann}, S.~G., {Temmer}, M., {et~al.} 2021, \apj, 918,
  21, \dodoi{10.3847/1538-4357/ac090a}

\bibitem[{{Lionello} {et~al.}(2001){Lionello}, {Linker}, \&
  {Miki{\'c}}}]{lionello01}
{Lionello}, R., {Linker}, J.~A., \& {Miki{\'c}}, Z. 2001, \apj, 546, 542,
  \dodoi{10.1086/318254}

\bibitem[{{Lionello} {et~al.}(2009){Lionello}, {Linker}, \&
  {Miki{\'c}}}]{lionelloetal2009}
---. 2009, \apj, 690, 902, \dodoi{10.1088/0004-637X/690/1/902}

\bibitem[{{Lionello} {et~al.}(1998){Lionello}, {Mikic}, \&
  {Schnack}}]{lionelloetal1998}
{Lionello}, R., {Mikic}, Z., \& {Schnack}, D.~D. 1998, Journal of Computational
  Physics, 140, 172, \dodoi{10.1006/jcph.1998.5841}

\bibitem[{{Lockwood} {et~al.}(2009{\natexlab{a}}){Lockwood}, {Owens}, \&
  {Rouillard}}]{Lockwood_1_2009}
{Lockwood}, M., {Owens}, M., \& {Rouillard}, A.~P. 2009{\natexlab{a}}, Journal
  of Geophysical Research (Space Physics), 114, A11103,
  \dodoi{10.1029/2009JA014449}

\bibitem[{{Lockwood} {et~al.}(2009{\natexlab{b}}){Lockwood}, {Owens}, \&
  {Rouillard}}]{Lockwood_2_2009}
---. 2009{\natexlab{b}}, Journal of Geophysical Research (Space Physics), 114,
  A11104, \dodoi{10.1029/2009JA014450}

\bibitem[{{Manoharan} {et~al.}(2015){Manoharan}, {Kim}, {Pogorelov}, {Arge}, \&
  {Manoharan}}]{manoharan15}
{Manoharan}, P., {Kim}, T., {Pogorelov}, N.~V., {Arge}, C.~N., \& {Manoharan},
  P.~K. 2015, in Journal of Physics Conference Series, Vol. 642, Journal of
  Physics Conference Series, 012016, \dodoi{10.1088/1742-6596/642/1/012016}

\bibitem[{{McGregor} {et~al.}(2008){McGregor}, {Hughes}, {Arge}, \&
  {Owens}}]{mcgregor08}
{McGregor}, S.~L., {Hughes}, W.~J., {Arge}, C.~N., \& {Owens}, M.~J. 2008,
  Journal of Geophysical Research (Space Physics), 113, A08112,
  \dodoi{10.1029/2007JA012330}

\bibitem[{{McGregor} {et~al.}(2011){McGregor}, {Hughes}, {Arge}, {Owens}, \&
  {Odstrcil}}]{mcgregor11}
{McGregor}, S.~L., {Hughes}, W.~J., {Arge}, C.~N., {Owens}, M.~J., \&
  {Odstrcil}, D. 2011, Journal of Geophysical Research (Space Physics), 116,
  A03101, \dodoi{10.1029/2010JA015881}

\bibitem[{{Merkin} {et~al.}(2016){Merkin}, {Lyon}, {Lario}, {Arge}, \&
  {Henney}}]{merkin16}
{Merkin}, V.~G., {Lyon}, J.~G., {Lario}, D., {Arge}, C.~N., \& {Henney}, C.~J.
  2016, Journal of Geophysical Research (Space Physics), 121, 2866,
  \dodoi{10.1002/2015JA022200}

\bibitem[{{Mikic} \& {Linker}(1994)}]{mikiclinker94}
{Mikic}, Z., \& {Linker}, J.~A. 1994, \apj, 430, 898, \dodoi{10.1086/174460}

\bibitem[{{Miki{\'c}} {et~al.}(1999){Miki{\'c}}, {Linker}, {Schnack},
  {Lionello}, \& {Tarditi}}]{mikicetal99}
{Miki{\'c}}, Z., {Linker}, J.~A., {Schnack}, D.~D., {Lionello}, R., \&
  {Tarditi}, A. 1999, Physics of Plasmas, 6, 2217, \dodoi{10.1063/1.873474}

\bibitem[{{Miki{\'c}} {et~al.}(2018){Miki{\'c}}, {Downs}, {Linker}, {Caplan},
  {Mackay}, {Upton}, {Riley}, {Lionello}, {T{\"o}r{\"o}k}, {Titov}, {Wijaya},
  {Druckm{\"u}ller}, {Pasachoff}, \& {Carlos}}]{mikicetal2018}
{Miki{\'c}}, Z., {Downs}, C., {Linker}, J.~A., {et~al.} 2018, Nature Astronomy,
  2, 913, \dodoi{10.1038/s41550-018-0562-5}

\bibitem[{{Odstrcil}(2003)}]{odstrcil03}
{Odstrcil}, D. 2003, Advances in Space Research, 32, 497,
  \dodoi{10.1016/S0273-1177(03)00332-6}

\bibitem[{{Odstrcil} {et~al.}(2005){Odstrcil}, {Pizzo}, \& {Arge}}]{odstrcil05}
{Odstrcil}, D., {Pizzo}, V.~J., \& {Arge}, C.~N. 2005, Journal of Geophysical
  Research (Space Physics), 110, A02106, \dodoi{10.1029/2004JA010745}

\bibitem[{{Owens} {et~al.}(2018){Owens}, {Lockwood}, {Barnard}, \&
  {MacNeil}}]{Owens_2018}
{Owens}, M.~J., {Lockwood}, M., {Barnard}, L.~A., \& {MacNeil}, A.~R. 2018,
  \apjl, 868, L14, \dodoi{10.3847/2041-8213/aaee82}

\bibitem[{{Pahud} {et~al.}(2012){Pahud}, {Merkin}, {Arge}, {Hughes}, \&
  {McGregor}}]{pahud12}
{Pahud}, D.~M., {Merkin}, V.~G., {Arge}, C.~N., {Hughes}, W.~J., \& {McGregor},
  S.~M. 2012, Journal of Atmospheric and Solar-Terrestrial Physics, 83, 32,
  \dodoi{10.1016/j.jastp.2012.02.012}

\bibitem[{{Panasenco} {et~al.}(2020){Panasenco}, {Velli}, {D'Amicis}, {Shi},
  {R{\'e}ville}, {Bale}, {Badman}, {Kasper}, {Korreck}, {Bonnell}, {Wit},
  {Goetz}, {Harvey}, {MacDowall}, {Malaspina}, {Pulupa}, {Case}, {Larson},
  {Livi}, {Stevens}, \& {Whittlesey}}]{panasenco20}
{Panasenco}, O., {Velli}, M., {D'Amicis}, R., {et~al.} 2020, \apjs, 246, 54,
  \dodoi{10.3847/1538-4365/ab61f4}

\bibitem[{{Perri} {et~al.}(2022){Perri}, {Leitner}, {Brchnelova},
  {Baratashvili}, {Kuzma}, {Zhang}, {Lani}, \& {Poedts}}]{perri2022}
{Perri}, B., {Leitner}, P., {Brchnelova}, M., {et~al.} 2022, arXiv e-prints,
  arXiv:2205.03341.
\newblock \doarXiv{2205.03341}

\bibitem[{{Pinto} \& {Rouillard}(2017)}]{pinto17}
{Pinto}, R.~F., \& {Rouillard}, A.~P. 2017, \apj, 838, 89,
  \dodoi{10.3847/1538-4357/aa6398}

\bibitem[{{Pomoell} \& {Poedts}(2018)}]{pomoell18}
{Pomoell}, J., \& {Poedts}, S. 2018, Journal of Space Weather and Space
  Climate, 8, A35, \dodoi{10.1051/swsc/2018020}

\bibitem[{{Prabhu} {et~al.}(2020){Prabhu}, {Lagg}, {Hirzberger}, \&
  {Solanki}}]{prabhuetal20}
{Prabhu}, A., {Lagg}, A., {Hirzberger}, J., \& {Solanki}, S.~K. 2020, \aap,
  644, A86, \dodoi{10.1051/0004-6361/202038704}

\bibitem[{{Riley} {et~al.}(2001){Riley}, {Linker}, \&
  {Miki{\'c}}}]{rileyetal01}
{Riley}, P., {Linker}, J.~A., \& {Miki{\'c}}, Z. 2001, \jgr, 106, 15889,
  \dodoi{10.1029/2000JA000121}

\bibitem[{{Riley} {et~al.}(2019){Riley}, {Linker}, {Mikic}, {Caplan}, {Downs},
  \& {Thumm}}]{rileyetal19}
{Riley}, P., {Linker}, J.~A., {Mikic}, Z., {et~al.} 2019, \apj, 884, 18,
  \dodoi{10.3847/1538-4357/ab3a98}

\bibitem[{{Riley} {et~al.}(2006){Riley}, {Linker}, {Miki{\'c}}, {Lionello},
  {Ledvina}, \& {Luhmann}}]{rileyetal06}
{Riley}, P., {Linker}, J.~A., {Miki{\'c}}, Z., {et~al.} 2006, \apj, 653, 1510,
  \dodoi{10.1086/508565}

\bibitem[{{Riley} {et~al.}(2014){Riley}, {Ben-Nun}, {Linker}, {Mikic},
  {Svalgaard}, {Harvey}, {Bertello}, {Hoeksema}, {Liu}, \&
  {Ulrich}}]{riley2014}
{Riley}, P., {Ben-Nun}, M., {Linker}, J.~A., {et~al.} 2014, \solphys, 289, 769,
  \dodoi{10.1007/s11207-013-0353-1}

\bibitem[{{Sachdeva} {et~al.}(2019){Sachdeva}, {van der Holst}, {Manchester},
  {T{\'o}th}, {Chen}, {Lloveras}, {V{\'a}squez}, {Lamy}, {Wojak}, {Jackson},
  {Yu}, \& {Henney}}]{sachdevaetal19}
{Sachdeva}, N., {van der Holst}, B., {Manchester}, W.~B., {et~al.} 2019, \apj,
  887, 83, \dodoi{10.3847/1538-4357/ab4f5e}

\bibitem[{{Sachdeva} {et~al.}(2021){Sachdeva}, {T{\'o}th}, {Manchester}, {van
  der Holst}, {Huang}, {Sokolov}, {Zhao}, {Shidi}, {Chen}, {Gombosi}, {Henney},
  {Lloveras}, \& {V{\'a}squez}}]{sachdevaetal21}
{Sachdeva}, N., {T{\'o}th}, G., {Manchester}, W.~B., {et~al.} 2021, \apj, 923,
  176, \dodoi{10.3847/1538-4357/ac307c}

\bibitem[{{Schatten}(1971)}]{schatten71}
{Schatten}, K.~H. 1971, Cosmic Electrodynamics, 2, 232

\bibitem[{{Schatten} {et~al.}(1969){Schatten}, {Wilcox}, \&
  {Ness}}]{schatten69}
{Schatten}, K.~H., {Wilcox}, J.~M., \& {Ness}, N.~F. 1969, \solphys, 6, 442,
  \dodoi{10.1007/BF00146478}

\bibitem[{{Schonfeld} {et~al.}(2022){Schonfeld}, {Henney}, {Jones}, \&
  {Arge}}]{schonfeldetal22}
{Schonfeld}, S.~J., {Henney}, C.~J., {Jones}, S.~I., \& {Arge}, C.~N. 2022,
  \apj, 932, 115, \dodoi{10.3847/1538-4357/ac6ba1}

\bibitem[{{Schou} {et~al.}(2012){Schou}, {Scherrer}, {Bush}, {Wachter},
  {Couvidat}, {Rabello-Soares}, {Bogart}, {Hoeksema}, {Liu}, {Duvall}, {Akin},
  {Allard}, {Miles}, {Rairden}, {Shine}, {Tarbell}, {Title}, {Wolfson},
  {Elmore}, {Norton}, \& {Tomczyk}}]{Schou2012}
{Schou}, J., {Scherrer}, P.~H., {Bush}, R.~I., {et~al.} 2012, \solphys, 275,
  229, \dodoi{10.1007/s11207-011-9842-2}

\bibitem[{{Schwadron} {et~al.}(2005){Schwadron}, {McComas}, {Elliott},
  {Gloeckler}, {Geiss}, \& {von Steiger}}]{schwadron2005}
{Schwadron}, N.~A., {McComas}, D.~J., {Elliott}, H.~A., {et~al.} 2005, Journal
  of Geophysical Research (Space Physics), 110, A04104,
  \dodoi{10.1029/2004JA010896}

\bibitem[{{Schwenn}(2006{\natexlab{a}})}]{schwenn06a}
{Schwenn}, R. 2006{\natexlab{a}}, \ssr, 124, 51,
  \dodoi{10.1007/s11214-006-9099-5}

\bibitem[{{Schwenn}(2006{\natexlab{b}})}]{schwenn06b}
---. 2006{\natexlab{b}}, Living Reviews in Solar Physics, 3, 2,
  \dodoi{10.12942/lrsp-2006-2}

\bibitem[{{Singh} {et~al.}(2020){Singh}, {Kim}, {Pogorelov}, \&
  {Arge}}]{singh20}
{Singh}, T., {Kim}, T.~K., {Pogorelov}, N.~V., \& {Arge}, C.~N. 2020, Space
  Weather, 18, e02405, \dodoi{10.1029/2019SW002405}

\bibitem[{{Solanki} {et~al.}(2020){Solanki}, {del Toro Iniesta}, {Woch},
  {Gandorfer}, {Hirzberger}, {Alvarez-Herrero}, {Appourchaux}, {Mart{\'\i}nez
  Pillet}, {P{\'e}rez-Grande}, {Sanchis Kilders}, {Schmidt}, {G{\'o}mez Cama},
  {Michalik}, {Deutsch}, {Fernandez-Rico}, {Grauf}, {Gizon}, {Heerlein},
  {Kolleck}, {Lagg}, {Meller}, {M{\"u}ller}, {Sch{\"u}hle}, {Staub}, {Albert},
  {Alvarez Copano}, {Beckmann}, {Bischoff}, {Busse}, {Enge}, {Frahm},
  {Germerott}, {Guerrero}, {L{\"o}ptien}, {Meierdierks}, {Oberdorfer},
  {Papagiannaki}, {Ramanath}, {Schou}, {Werner}, {Yang}, {Zerr}, {Bergmann},
  {Bochmann}, {Heinrichs}, {Meyer}, {Monecke}, {M{\"u}ller}, {Sperling},
  {{\'A}lvarez Garc{\'\i}a}, {Aparicio}, {Balaguer Jim{\'e}nez}, {Bellot
  Rubio}, {Cobos Carracosa}, {Girela}, {Hern{\'a}ndez Exp{\'o}sito}, {Herranz},
  {Labrousse}, {L{\'o}pez Jim{\'e}nez}, {Orozco Su{\'a}rez}, {Ramos},
  {Barandiar{\'a}n}, {Bastide}, {Campuzano}, {Cebollero}, {D{\'a}vila},
  {Fern{\'a}ndez-Medina}, {Garc{\'\i}a Parejo}, {Garranzo-Garc{\'\i}a},
  {Laguna}, {Mart{\'\i}n}, {Navarro}, {N{\'u}{\~n}ez Peral}, {Royo},
  {S{\'a}nchez}, {Silva-L{\'o}pez}, {Vera}, {Villanueva}, {Fourmond}, {de
  Galarreta}, {Bouzit}, {Hervier}, {Le Clec'h}, {Szwec}, {Chaigneau},
  {Buttice}, {Dominguez-Tagle}, {Philippon}, {Boumier}, {Le Cocguen},
  {Baranjuk}, {Bell}, {Berkefeld}, {Baumgartner}, {Heidecke}, {Maue}, {Nakai},
  {Scheiffelen}, {Sigwarth}, {Soltau}, {Volkmer}, {Blanco Rodr{\'\i}guez},
  {Domingo}, {Ferreres Sabater}, {Gasent Blesa}, {Rodr{\'\i}guez
  Mart{\'\i}nez}, {Osorno Caudel}, {Bosch}, {Casas}, {Carmona}, {Herms},
  {Roma}, {Alonso}, {G{\'o}mez-Sanjuan}, {Piqueras}, {Torralbo}, {Fiethe},
  {Guan}, {Lange}, {Michel}, {Bonet}, {Fahmy}, {M{\"u}ller}, \&
  {Zouganelis}}]{solankietal20}
{Solanki}, S.~K., {del Toro Iniesta}, J.~C., {Woch}, J., {et~al.} 2020, \aap,
  642, A11, \dodoi{10.1051/0004-6361/201935325}

\bibitem[{{Sun} {et~al.}(2011){Sun}, {Liu}, {Hoeksema}, {Hayashi}, \&
  {Zhao}}]{sunetal11}
{Sun}, X., {Liu}, Y., {Hoeksema}, J.~T., {Hayashi}, K., \& {Zhao}, X. 2011,
  \solphys, 270, 9, \dodoi{10.1007/s11207-011-9751-4}

\bibitem[{{van der Holst} {et~al.}(2010){van der Holst}, {Manchester},
  {Frazin}, {V{\'a}squez}, {T{\'o}th}, \& {Gombosi}}]{vanderholstetal10}
{van der Holst}, B., {Manchester}, W.~B., I., {Frazin}, R.~A., {et~al.} 2010,
  \apj, 725, 1373, \dodoi{10.1088/0004-637X/725/1/1373}

\bibitem[{{van der Holst} {et~al.}(2014){van der Holst}, {Sokolov}, {Meng},
  {Jin}, {Manchester}, {T{\'o}th}, \& {Gombosi}}]{vanderholstetal14}
{van der Holst}, B., {Sokolov}, I.~V., {Meng}, X., {et~al.} 2014, \apj, 782,
  81, \dodoi{10.1088/0004-637X/782/2/81}

\bibitem[{{Wagner} {et~al.}(2022){Wagner}, {Asvestari}, {Temmer}, {Heinemann},
  \& {Pomoell}}]{wagner22}
{Wagner}, A., {Asvestari}, E., {Temmer}, M., {Heinemann}, S.~G., \& {Pomoell},
  J. 2022, \aap, 657, A117, \dodoi{10.1051/0004-6361/202141552}

\bibitem[{{Wallace} {et~al.}(2019){Wallace}, {Arge}, {Pattichis},
  {Hock-Mysliwiec}, \& {Henney}}]{wallace19}
{Wallace}, S., {Arge}, C.~N., {Pattichis}, M., {Hock-Mysliwiec}, R.~A., \&
  {Henney}, C.~J. 2019, \solphys, 294, 19, \dodoi{10.1007/s11207-019-1402-1}

\bibitem[{{Wang} {et~al.}(2020){Wang}, {Chen}, {Hu}, {Jiang}, {Song}, {Wu}, \&
  {Ning}}]{wang_etal20}
{Wang}, B., {Chen}, Y., {Hu}, Q., {et~al.} 2020, Science in China E:
  Technological Sciences, 63, 234, \dodoi{10.1007/s11431-018-9470-y}

\bibitem[{{Wang} {et~al.}(1996){Wang}, {Hawley}, \& {Sheeley}}]{wang96}
{Wang}, Y.-M., {Hawley}, S.~H., \& {Sheeley}, Neil~R., J. 1996, Science, 271,
  464, \dodoi{10.1126/science.271.5248.464}

\bibitem[{{Wang} \& {Sheeley}(1992)}]{wang92}
{Wang}, Y.~M., \& {Sheeley}, N.~R., J. 1992, \apj, 392, 310,
  \dodoi{10.1086/171430}

\bibitem[{{Wang} \& {Sheeley}(1995)}]{wang95}
---. 1995, \apjl, 447, L143, \dodoi{10.1086/309578}

\bibitem[{{Wang} {et~al.}(2022){Wang}, {Ulrich}, \& {Harvey}}]{wang22}
{Wang}, Y.~M., {Ulrich}, R.~K., \& {Harvey}, J.~W. 2022, \apj, 926, 113,
  \dodoi{10.3847/1538-4357/ac4491}

\bibitem[{{Wiegelmann}(2007)}]{wiegelmann_etal07}
{Wiegelmann}, T. 2007, \solphys, 240, 227, \dodoi{10.1007/s11207-006-0266-3}

\bibitem[{{Zhang} {et~al.}(2019){Zhang}, {Sorathia}, {Lyon}, {Merkin},
  {Garretson}, \& {Wiltberger}}]{zhang19}
{Zhang}, B., {Sorathia}, K.~A., {Lyon}, J.~G., {et~al.} 2019, \apjs, 244, 20,
  \dodoi{10.3847/1538-4365/ab3a4c}

\end{thebibliography}
\bibliographystyle{aasjournal}



\end{document}